\documentclass[12pt]{iopart}
\usepackage[english]{babel}
\usepackage[utf8]{inputenc}
\usepackage{dcolumn}    
\usepackage{bm} 
\usepackage{graphicx}
\usepackage{enumitem}
\usepackage{latexsym}
\usepackage{array}      
\usepackage{epsfig}
\usepackage{braket}
\usepackage{color}
\usepackage[colorlinks=true,linkcolor=blue,urlcolor=blue,citecolor=blue,pdfusetitle]{hyperref}
\usepackage{hyperref}
\usepackage[T1]{fontenc}
\usepackage[sc]{mathpazo}
\graphicspath{{./figs/}}

\newcommand{\ignore}[1]{}

\newcommand{\eq}{Eq.\,}
\newcommand{\eqs}{Eqs.\,}
\newcommand{\fig}{Fig.\,}
\newcommand{\figs}{Figs.\,}
\newcommand{\cf} {cf.~}

\newcommand{\ie} {i.e.~}
\newcommand{\eg} {e.g.~}
\newcommand{\rref} {Ref.\,}
\newcommand{\rrefs} {Refs.\,}

\newcommand{\Hc} {\rm {H.c.}}
\newcommand{\NN} {\mathcal{N}}
\newcommand{\ombs} {\omega_{\rm BS}}

\newcommand{\bfk} {\bf k}
\newcommand{\bfr} {\bf r}

\newcommand{\eqref}[1]{(\ref{#1})}

\begin{document}

	\title{Quantum optics with giant atoms in a structured photonic bath}

	\author{Luca Leonforte$^{1,2,*}$, Xuejian Sun$^{3,4}$, Davide Valenti$^5$, Bernardo Spagnolo$^5$, Fabrizio Illuminati$^{2,6,8}$, Angelo Carollo$^3$, Francesco Ciccarello$^{3,7}$}
	
	\address{$^1$Dipartimento di Fisica "E.R. Caianello", Università degli Studi di Salerno, Via Giovanni Paolo II, 132 I-84084 Fisciano (SA), Italy}
	\address{$^2$INFN, Sezione di Napoli, Gruppo collegato di Salerno, Italy.}
	\address{$^3$Università degli Studi di Palermo, Dipartimento di Fisica e Chimica -- Emilio Segrè, via Archirafi 36, I-90123 Palermo, Italy}
	\address{$^4$School of Physics and Telecommunication Engineering, Zhoukou Normal University, Zhoukou, 466001, China}
	\address{$^5$Dipartimento di Fisica e Chimica "Emilio Segrè", Group of Interdisciplinary Theoretical Physics, Università di Palermo, Viale delle Scienze, Ed.18, I-90128 Palermo, Italy}
	\address{$^6$Dipartimento di Ingegneria Industriale, Università degli Studi di Salerno, Via Giovanni Paolo II, 132 I-84084 Fisciano (SA), Italy}
	\address{$^7$NEST, Istituto Nanoscienze-CNR, Piazza S. Silvestro 12, 56127 Pisa, Italy}
	\address{$^8$Institute of Nanotechnology, CNR NANOTEC, Via Monteroni, 73100 Lecce, Italy}
    \address{$^*$Author to whom any correspondence should be addressed}
	\eads{luca.leonforte93@gmail.com}
	%	\begin{indented}
		%		\item[]\today
		%	\end{indented}
	
	\begin{abstract}
		We present a general framework to tackle quantum optics problems with giant atoms, \ie quantum emitters each coupled {\it non-locally} to a structured photonic bath (typically a lattice) of any dimension. The theory encompasses the calculation and general properties of Green's functions, atom-photon bound states (BSs), collective master equations and decoherence-free Hamiltonians (DFHs), and is underpinned by a formalism where a giant atom is formally viewed as a normal atom lying at a fictitious location. As a major application, we provide for the first time a general  criterion to predict/engineer DFHs of giant atoms, which can be applied both in and out of the photonic continuum and regardless of the structure or dimensionality of the photonic bath. This is used to show novel DFHs in 2D baths such as a square lattice,  photonic graphene and an extended photonic Lieb lattice.	
	\end{abstract}
	
	\vspace{2pc}
	
	\submitto{Quantum Sci. Technol.}		% cannot find the proper command of the journal
	\maketitle

	\section{Introduction}

\begin{figure}[t!]
		\begin{center}
			\includegraphics[width=0.75\textwidth]{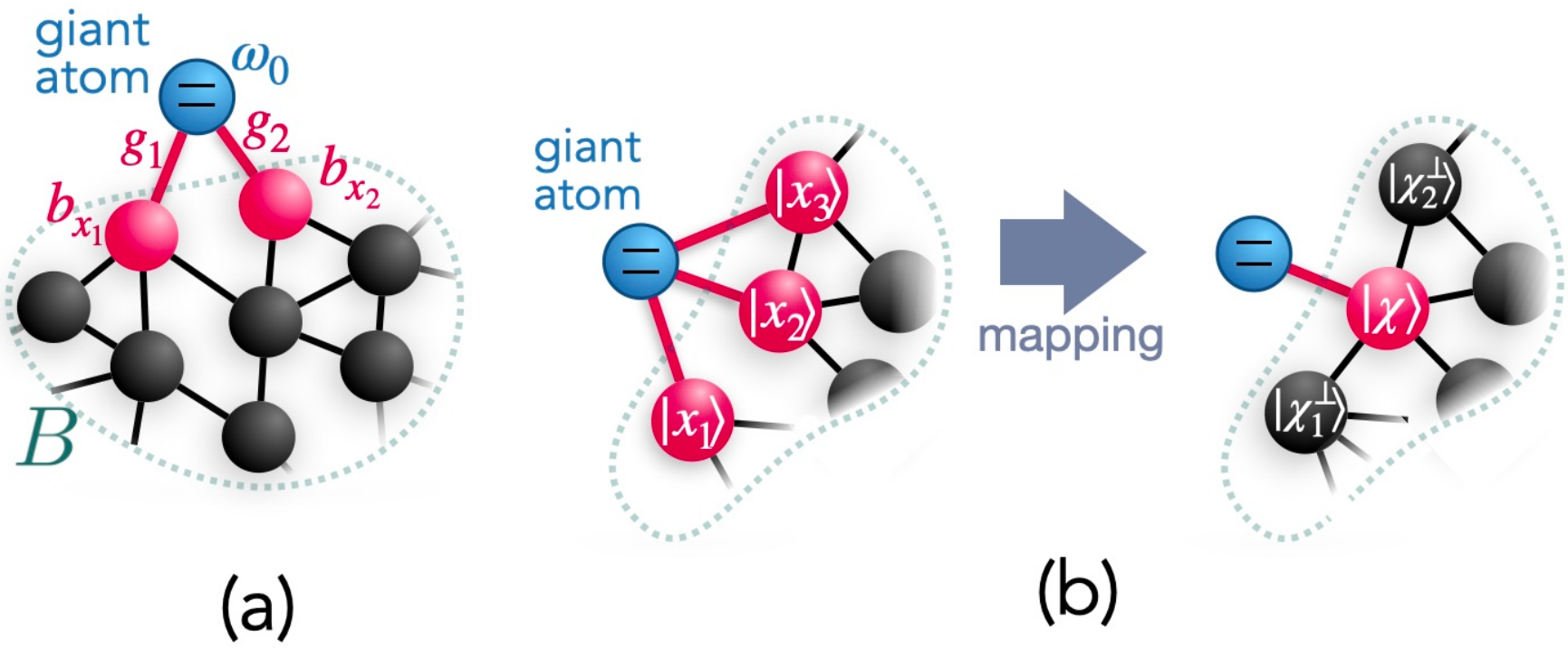}
			\caption{(a) Giant atom (two-level system of frequency $\omega_{0}$) non-locally coupled to a photonic bath $B$ modeled as a set of coupled cavities. There are only two coupling points in this example with coupling strengths $g_1$ and $g_2$. (b) Under a suitable mapping (unitary $U_\chi$ on $B$) the giant atom can be transformed into a normal one featuring a single coupling point corresponding to the single-photon state $\ket{\chi}$ ("site state").} \label{fig-setup}
		\end{center}
	\end{figure}

	In recent years, it became experimentally possible realizing so called {\it giant atoms} \cite{Kockum5years}. A giant atom is a (usually artificial) quantum emitter which interacts coherently with the field at a discrete set of coupling points [see \fig\ref{fig-setup}(a)]. Such a {\it non-local} coupling was first achieved in circuit QED (see \eg \rrefs\cite{gu2017microwave,blais2021circuit}) for superconducting qubits coupled to a 1D waveguide along which either phonons \cite{gustafsson2014propagating} or microwave photons \cite{kannan2020waveguide,vadiraj2021engineering} can propagate, where the distance between coupling points can be made comparable with the carrier wavelength, thus enabling unprecedented self-interference effects. A recent experiment used a ferromagnetic spin ensemble coupled to a meandering waveguide \cite{wang2022giant}, while implementations based on ultracold atoms \cite{gonzalez2019engineering} and Rydberg atoms in photonic crystal waveguides \cite{chen2023giant} were recently put forward. Notably, the coupling of a giant atom to the field at each point can be made complex with a controllable phase, which was recently demonstrated \cite{joshi2023resonance}.

	Giant atoms allow for quantum optics phenomena which are impossible with normal atoms (\ie with the standard local coupling). A remarkable effect, predicted theoretically \cite{KockumPRL2018} and experimentally confirmed in a circuit-QED setup \cite{kannan2020waveguide}, is occurrence of {\it decoherence-free Hamiltonians} (DFHs) between atoms mediated by the field of a 1D waveguide \cite{karg2019remote,DarioTB,carollo2020mechanism,soro2022chiral}. These are purely dispersive dipole-dipole interactions, described by an effective many-body spin Hamiltonian, which arise for suitable arrangements of the coupling points when the atomic frequency lies {\it in} the field's continuum. This is achieved through destructive interference which fully suppresses dissipation into the photonic bath, a task out of reach with normal (\ie point-like) atoms. Moreover, in some settings such as acoustic waveguides, retardation times associated with the coupling points' distance can be made long compared to the atom decay time resulting in unprecedented non-Markovian phenomena \cite{GuoPRA17,andersson2019non,lim2023oscillating,KockumPRR20,qiu2023collective}. Recently, the study of giant atoms was extended to the ultrastrong coupling regime \cite{terradas2022ultrastrong,noachtar2022nonperturbative}.

	While there exists already a significant body of literature on giant atoms in continuous waveguides, their study in discrete {\it structured} baths and especially photonic lattices is still in the early stages \cite{zhao2020single,VegaPRA21,wang2021tunable,yu2021entanglement,cheng2022topology,RongPRA22,xiao2022bound,soro2023interaction,DuQST2023,wang2023giant,dong2023exotic,Zhang2023quantum,wang2023giant,vega2023topological,du2023giant}.  In particular, it was predicted formation of atom-photon BSs and occurrence of DFHs, both in and out of the continuum, in some specific coupled-cavity arrays. To the best of our knowledge, however, no general study of giant-atom properties in a {\it model-independent} fashion was carried out so far, \ie without considering a specific photonic bath structure, which appears an indispensable step to devise general criteria for engineering future giant-atoms setups.
	
	Another issue of methodological relevance, but which will turn out to have conceptual implications, is the following.
	The most common approach adopted so far to investigate giant atoms in structured baths is to apply the resolvent (\ie Green's-function) method \cite{Kofman1994,Lambropoulos2000a} to the case of multiple coupling points and use the bath normal modes, which leads to a picture where the giant atom can be formally viewed as a normal atom but with a modified coupling strength to each bath normal mode. One can ask whether some sort of mapping to a normal (point-like) atom is possible even in {\it real space} and, if so, whether it brings any advantage. While this entails dealing with local bath modes in place of normal ones, real-space approaches proved powerful in a number of quantum optics problems (\eg atom-photon scattering in waveguides). We will show that this is the case even with giant atoms, allowing \eg to bypass dealing explicitly with the Green's function in certain problems.
	
	A further issue concerns DFHs: for normal atoms (local coupling) these are well-known to be mediated by {\it atom-photon bound states} (BSs) out of the continuum, typically within bandgaps of photonic lattices \cite{Lambropoulos2000a,Calajo2016b,Liu2017a,Krinner2018,DominikPRR20} and are investigated both theoretically \cite{Douglas2015b,gonzalez2015subwavelength,hood2016atom, shi2018effective,gonzalez2018exotic,Bello2019,LeonfortePRL2021,tabares2023variational} and experimentally \cite{Sundaresan2019,painter205,ScigliuzzoPRX22}. It is natural to ask whether an analogous interpretation in terms of BSs can be made for DFHs of giant atoms, but in terms of BSs {\it in} the photonic continuum (which cannot occur for normal atoms). 
	This issue is far from being a merely theoretical curiosity since (as will become clear later) it is deeply connected with the identification of a model- and dimension-independent physical mechanism behind occurrence of DFHs, a task that has not been carried out to date.

	With the above motivations, in this paper we present a general theory of giant atoms in structured photonic baths (having in mind mostly waveguides and photonic lattices of arbitrary dimension, even higher than 1D). Despite the coupling is non-local, we first arrange the atom-field interaction Hamiltonian in a way formally analogous to a normal atom by defining a fictitious giant-atom's location (and a related single-photon state), which depends on the pattern of coupling points and their interaction strengths. This is used to derive a general and very compact expression of the atom-photon Green's function (resolvent) in terms of the bare bath Green's function, which can be exploited to derive a number of quantum optics properties. We apply this resolvent formalism to establish some general features and occurrence conditions of atom-photon BSs of giant atoms and to work out a general collective Lindblad master equation. This describes the  open dynamics of a set of giant atoms in a structured zero-temperature bath, the rates of which are each expressed (taking advantage of the fictitious giant-atom locations) in terms of two-point matrix elements of the bath resolvent (thus formally like normal atoms). We make use of the master equation to prove that occurrence of a DFH Hamiltonian is equivalent to occurrence of one bound state for each giant atom (in or out of the continuum). When BSs have non-zero overlaps with fictitious position states of the giant atoms, a non-trivial DFH arises, which in particular explains in a very simple way why a braided configuration is required in a 1D waveguide. Most importantly, this BS picture is shown to be an effective tool for predicting new classes of DFHs (including higher dimensions), which we will illustrate through various examples after singling out a class of BSs called vacancy-like dressed states.
	
	Before proceeding further, we note that the giant atom definition which is conveniently adopted in this work -- namely a quantum emitter with more than one coupling points -- may not exactly match definitions used elsewehere (for which having multiple coupling points is certainly a necessary but in general not sufficient condition).

	This work is organized as follows. In Section \ref{sec-model}, we define the model and Hamiltonian and show how one can rearrange the atom-field interaction Hamiltonian as if the giant atom were a normal one, defining at the same time a fictitious location (and related single-photon state). 
	We continue in Section \ref{sec-green}  by deriving a compact expression for the general single-excitation Green's function of the joint system in terms of the fictitious position of the giant atom and the bare field's resolvent. As a major application, this is used in the following Section \ref{Bs-sec} to derive general properties of atom-photon bound states, both in and out of the continuum. In Section \ref{sec-ME}, we generalize the Hamiltonian model to the case of many giant atoms and present a general Lindblad master equation governing their open dynamics. This is then applied in the next Section \ref{sec-DFH} to formulate a general condition for occurrence of DFHs in terms of BSs and to arrange the effective atom-atom coupling strengths (defining the same Hamiltonian) in terms of overlapping BSs. In Section \ref{sec-vds}, we study an important class of BSs: vacancy-like dressed states and several instances are illustrated in the following Section \ref{sec-ex}. The study of such states, as shown in Section \ref{sec-VDS-DFH}, sheds new light on known DFHs and, most importantly, allows to predict new ones. As the 2D examples in Sections \ref{sec-ex} and \ref{sec-VDS-DFH} feature atoms with no less than three coupling points, one can wonder whether this is the minimum number to obtain BSs in the continuum in a 2D lattice. We show that this is not the case by presenting a counterexample in Section \ref{sec-2cps}, where VDSs and DFHs in the continuum arise with only two coupling points per atom. We conclude with a summary and discussion of the results in Section \ref{sec-conc}.

	\section{Model and Hamiltonian}\label{sec-model}

	We consider a general setup comprising a generic photonic bath $B$ and one giant atom, as sketched in \fig\ref{fig-setup}(a). Bath $B$ is modeled as a discrete network of single-mode coupled cavities each labeled by $x$ (standing for an integer or a set of integers) with $b_x$ the usual bosonic ladder operator destroying a photon in cavity $x$. It is worth noticing that this general model encompasses the standard continuous waveguide with linear dispersion law that is routinely considered in most works on giant atoms, which is indeed retrieved when $B$ is a 1D array of cavities weakly coupled to an atom whose frequency falls within a photonic band (allowing to linearize the photon dispersion law).
	
	The $B$'s free Hamiltonian has the general form
	\begin{equation}
		H_B=\sum_{x} \omega_x  b_x^\dag b_x+\!\sum_{x\neq x'} \,J_{xx'}  b_x^\dag b_{x'}\,\label{HB}
	\end{equation}
	with $\omega_x$ the frequency of cavity $x$ and $J_{xx'}$ the photon hopping rate between cavities $x$ and $x'$ \footnote{%\red
    {We assume each cavity $x$ to be a single-mode cavity, which requires the hopping rates $J_{xx’}$ to be much smaller than the frequency separation between cavity normal modes, an approximation usually well-matched in experiments.}}. 
	The giant atom is modeled as a two-level system of frequency $\omega_{0}$ and ground (excited) state $\ket{g}$ ($\ket{e}$), the corresponding ladder operators being $\sigma_{-}=\sigma_{+}^\dag=\ket{g}\!\bra{e}$ (the generalization to many giant atoms will be discussed later on in Section \ref{sec-ME}).  
	Assuming weak coupling, the photonic bath and giant atom interact according to the rotating-wave approximation. Importanly, due to the giant-atom nature of the quantum emitter, this interaction is generally {\it non-local}, meaning that the atom is directly coupled to ${\cal N}$ cavities of $B$ with ${\cal N}\ge 1$ (for $\NN=1$ we retrieve the standard case of a normal atom, \ie  local coupling).
	The total Hamiltonian thus reads
	\begin{equation}
		H=H_B+\omega_0 \sigma_+\sigma_-+\sum_{\ell=1}^{\cal N} \left(g_\ell \,b_{x_\ell}^\dag \sigma_-+\Hc\right)\label{H1}\,,
	\end{equation}
	where $g_\ell$ (generally complex) is the atom's coupling strength to cavity $x_\ell$ [see \fig\ref{fig-setup}(a)],  A cavity to which the atom is directly coupled (\ie cavity $x_\ell$ such that $g_\ell\neq 0$) will be often called "coupling point", their total number being ${\cal N}$.
	\\
	\\
	It is convenient to define the field ladder operator 
	\begin{equation}
		b_\chi=\sum_{\ell=1}^{\cal N} \alpha_\ell^* b_{x_\ell}\,\,\,{\rm with}\,\,\,\alpha_\ell=\frac{g_\ell}{\bar{g}}\,\,{\rm and}\,\,\bar{g}=\sqrt{\sum_\ell |g_\ell|^2}\,,\label{achi}
	\end{equation}
	which is a linear combination of the  ladder operators $b_{x_\ell}$ corresponding to the $\cal{N}$ coupling points $x_\ell$, and the coefficients $\alpha_\ell$’s are proportional to the related coupling strengths $g_\ell$’s.
	Since $\sum_{\ell=1}^{\cal N} |\alpha_\ell|^2=1$, $b_\chi$ fulfills $[b_\chi,b_\chi^\dag]=1$. With this definition the Hamiltonian \eqref{H1} can now be arranged as
	\begin{equation}
		H=H_B+\omega_0 \sigma_+\sigma_-+\bar{g} \left(b_{\chi}^\dag \sigma_-+\Hc\right)\label{H1-bis}\,,
	\end{equation}
	which is formally analogous to the case of a normal atom (notice that ${\bar{g}}>0$). We point out however that $b_{\chi}$ generally does not commute with field operators $b_{x}^\dag$, this being a signature of the non-local nature of atom-photon coupling.

	\subsection{Single-excitation sector}
	
	Due to the rotating-wave approximation the total number of excitations $\sum_x b_x^\dag b_x+\sigma_{+}\sigma_{-}$ is a constant of motion. The one-excitation sector is spanned by the set of states $\ket{e}\ket{\rm vac}$ and $\{\ket{g} \ket{x}\}$ with $\ket{\rm vac}$ the field's vacuum state and $\ket{x}=b_x^\dag\, |{\rm vac}\rangle$ the Fock state featuring one photon at cavity $x$. Since we will work mostly in this one-excitation subspace, it is convenient to adopt a light notation and replace 
	$$\ket{e}\ket{{\rm vac}}\rightarrow \ket{e}\,,\,\,\,\ket{g} \ket{x}\rightarrow \ket{x}\,\,.$$
	Thus $\ket{e}$ now denotes the state where one excitation lies on the atom (and there are no photons), while state $\ket{x}$ describes a single photon at cavity $x$ (with the atom in the ground state $\ket{g}$).
	
	The total Hamiltonian in this subspace can be conveniently arranged in the form
	\begin{equation}
		H^{(1)}=\omega_0|e\rangle\!\langle e|+H_B^{(1)}+\bar{g} \left( |\chi\rangle\!\langle e|+\Hc\right)\label{H2}
	\end{equation}
	where [\cf\eq\eqref{HB}]
	\begin{equation}
		H_B^{(1)}=\sum_{x=1}^N \omega_x  \ket{x}\!\bra{x}+\!\sum_{x\neq x'} \,J_{xx'}  \ket{x}\!\bra{x'}\label{HB2}\,
	\end{equation}
	is the field's free Hamiltonian in the one-excitation subspace, while 
	\begin{equation}
		\ket{\chi} =b_\chi^\dag |0\rangle= \sum_{\ell=1}^{\cal N} \alpha_\ell \ket{x_\ell}\,,\label{chi2}
	\end{equation}
	is a (normalized) single-photon state [recall \eq\eqref{achi}]. In the remainder, for convenience we will often refer to the single-photon state $\ket{\chi}$ as the {\it site state}.
	
	Henceforth, we will drop superscript "$(1)$" in both $H^{(1)}$ and $H_B^{(1)}$.

	\subsection{Mapping into a normal atom}\label{sec-uchi}
	
	A giant atom can be thought of as a normal atom which is yet coupled to a modified bath \cite{Kockum5years,gonzalez2019engineering}. For the discrete bath considered here, this can be easily seen in real space by replacing the ${\cal N}$ single-photon states $\{\ket{x_1}\,,\ket{x_2}\,,...,\ket{x_{\cal N}}\}$ (one for each coupling point) with $\{\ket{\chi^\perp_i},\ket{\chi}\}$, where $\ket{\chi}$ is the site state in [\cf\eq\eqref{chi2}] while $\ket{\chi^\perp_i}$ for $i=1,...,{\cal N}{-}1$ is any basis spanning the ($N$-1)-dimensional subspace of $\{\ket{x_1}\,,\ket{x_2}\,,...,\ket{x_{\cal N}}\}$ orthogonal to $\ket{\chi}$, \ie such that $\langle \chi^\perp_i|\chi\rangle=0$ and $\langle \chi^\perp_i|\chi^\perp_{i'}\rangle=\delta_{ii'}$ [see \fig\ref{fig-setup}(b)]. 
	\begin{figure}[t!]
		\begin{center}
			\includegraphics[width=0.75\textwidth]{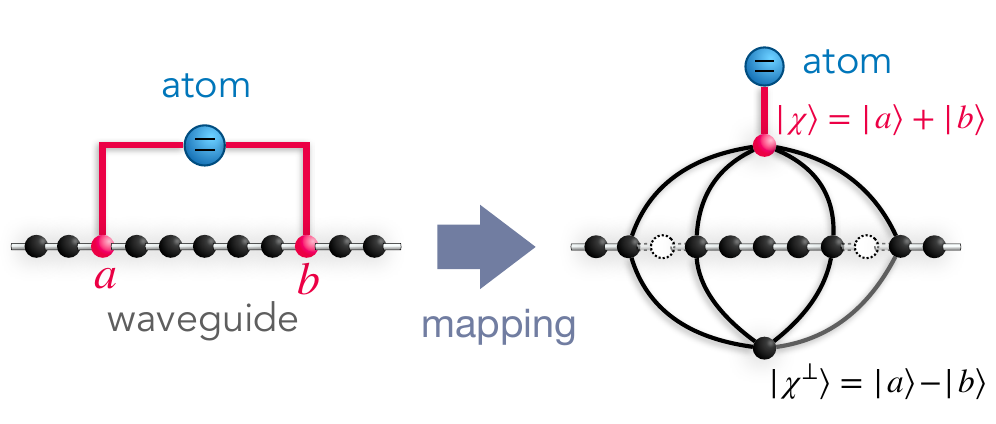}
			\caption{Left: giant atom with two coupling points (labeled $a$ and $b$) interacting with a coupled-cavity array (waveguide) having only nearest-neighbour hopping rates. For $g_a=g_b=g$, the site state $\ket{\chi}$ in this case is the symmetric superposition of $\ket{a}$ and $\ket{b}$. The mapping defined by $U_\chi$ turns the giant atom into a normal atom, but at the same time transforms the waveguide in a non-trivial way (right panel): cavities $a$ and $b$ are now replaced by their symmetric and antisymmetric combinations (each of these coupling now to four different cavities).} \label{fig-chi}
		\end{center}
	\end{figure}
	
	This corresponds to a unitary transformation $U_\chi$ on the field's Hilbert space such that the atom is now coupled only to the fictitious site corresponding to the site state $\ket{\chi}$  [see \fig\ref{fig-setup}(b)]. The cost of this mapping is that unitary $U_\chi$ will also change the bath Hamiltonian as $H_B\rightarrow H_B'=U_\chi H_B U_\chi^\dag $, where $H_B'$ may be more complicated to cope with compared to $H_B$, \eg because symmetries of the latter (such as translational symmetry) might no longer hold for $H_B'$. A simple but illustrative example is shown in \fig\ref{fig-chi} in the case that $B$ is a coupled-cavity array and an atom with two coupling points, showing that $H_B'$ breaks translational invariance. Despite this potential drawback, we will nevertheless see that this new picture, where the giant atom is turned into a normal atom (yet coupled to a modified bath), is useful to establish some general properties, e.g. concerning formation of atom-photon bound states.

	\section{Green's function}\label{sec-green}
	
	The Green's function (or resolvent operator) \cite{Economou2006} is an effective theoretical tool which is particularly suited to investigate atom-photon interactions \cite{CohenAP,Kofman1994,Lambropoulos2000a}.
	
	The Green's function associated with the total atom-field system is defined as $G(z)=1/(z-H)$ with $H$ the total Hamiltonian [\cf\eq\eqref{H1-bis}]. In \rref\cite{leonforte2021dressed}, it was shown that in the single-excitation sector $G(z)$ can be conveniently arranged in a compact form in terms of the bare field's Hamiltonian. The generalization of this expression to a giant atom is immediate thanks to \eq \eqref{H2} and reads \footnote{Indeed, to derive \eqref{Green} \cite{leonforte2021dressed} the only essential requirement is that the interaction Hamitonian has the form $\propto \!(\ket{\chi}\!\langle e|{+}{\rm H.c.})$ regardless of the nature of $\ket{\chi}$.}
	\begin{equation}
		\label{Green}
		G(z) = {G}_B ( z ) + \frac{1}{ F ( z ) }\, \ket{ \Psi(z)} \bra{ \Psi(z)} \,,
	\end{equation}
	where ${G}_B ( z )$ is the bath's Green function (more on this in the following subsection \ref{sec-TIB})~\footnote{There is a slight difference in the definition of $\ket{ \Psi(z)}$ and $F(z)$ used here and the one in \rref\cite{leonforte2021dressed}.}
	\begin{eqnarray}
		\ket{ \Psi(z)}& =&\ket{e} + \ket{\psi(z)}\,,\label{psiz}\\
		\ket{\psi(z)}&=&\bar{g} \,G_B ( z ) \ket{\chi}\,,\label{psiiz}\\
		F(z) & =& z-\omega_{0} -\bar{g}^2  \braket{\chi | {G}_B ( z ) | \chi}\label{fz}\,.
	\end{eqnarray}
	Like with normal atoms \cite{Lambropoulos2000a} the Green's function is helpful in particular for calculating atom-photon dressed states (both bound and unbound), providing at the same time the theoretical basis to establish important properties of these states. To accomplish such tasks, one can take advantage of the particularly compact expression \eqref{Green}. 
	
	Since \eqref{Green} is expressed in terms of the bath Green's function $G_B(z)$, in the next subsection we review the form of $G_B(z)$ in the common case that $B$ is a photonic lattice.

	\subsection{Translationally-invariant bath} \label{sec-TIB}
	
	When $B$ is a photonic lattice, discrete translational invariance holds. Then, based on the Bloch theorem the photonic energy spectrum consists of {\it bands} having dispersion law $\omega_{n\bfk}$ with $n$ the band index and wave vector ${\bf k}$ lying in the first Brillouin zone. The associated eigenstates are $\{\ket{\phi_{n{\bfk}}}\}$ such that $H_B \ket{\phi_{n{\bfk}}}=\omega_{n \bfk}\ket{\phi_{n\bfk}}$ with real-space wave function
	\begin{equation}
		\langle {\mathbf r},\beta \ket{\phi_{n{\mathbf k}}}=u_{n\beta}(\mathbf k)\,e^{i {\mathbf k}\cdot \mathbf r}\,,\label{phik1}
	\end{equation}
	where ${\bfr }$ is a discrete vector (belonging to a Bravais lattice) which identifies the unit cell, while $\beta$ is a discrete sublattice index (thus $\ket{x}$ in \eq\eqref{HB} is here embodied by $\ket{{\mathbf r},\beta}$).
	
	The bath Green's function therefore generally reads
	\begin{equation}
		G_B(z)=\sum_{n,\bfk} \frac{ \ket{\phi_{n{\bfk}}} \bra{\phi_{n{\bfk}}}}{z-\omega_{n\bfk}}\,.\label{GBB}
	\end{equation}
	If $\omega$ denotes a generic real frequency, it turns out that $G_B(\omega)$ is singular at any value of $\omega$ lying within an energy band [this being a branch cut of $G_B(z)$].  At such frequencies, $G_B(\omega)$ must be replaced with $ G_B(\omega^+)$, meaning that, in the denominator of \eqref{GBB} for $z{=}\omega$, $\omega$ must be turned into $\omega+i\epsilon$ with $\epsilon\rightarrow 0^+$. As a consequence, $ G_B(\omega^+)$ is generally complex inside bands. Such a singularity does not occur when $\omega$ lies in a photonic band gap, in which case $G_B(\omega)$ is real and coincides with $G_B(\omega^+)$.
	\\
	\\
	Although the Green's function is helpful also in the study of scattering (\ie unbound) states, in the remainder we will focus only on atom-photon {\it bound} states since they are key to the occurrence of decoherence-free Hamiltonians as will become clearer later on.

	\section{Atom-photon bound states} \label{Bs-sec}

	Atom-photon (dressed) bound states (BSs) have recently attracted a lot of interest, in particular because they can mediate atom-atom interactions with remarkable properties such as tunable interaction range and topological protection \cite{Douglas2015b,hood2016atom, gonzalez2018exotic,Bello2019,LeonfortePRL2021,Sundaresan2019,painter205,ScigliuzzoPRX22}.
	
	A dressed BS, $\ket{\Psi_{\rm BS}}$, is by definition a {\it normalizable} stationary state of the total system, that is 
	\begin{equation}
		H\ket{\Psi_{\rm BS}}=\omega_{\rm BS}\ket{\Psi_{\rm BS}}\,\,\,{\rm with}\,\,\,\langle \Psi_{\rm BS}\ket{\Psi_{\rm BS}}=1\,\,.\label{wbs}
	\end{equation}
	
	As is well-known, the poles of the Green's function are in one-to-one correspondence with bound eigenstates of $H$ \cite{Economou2006}, namely $z=\ombs$ with $\omega_{ \rm BS}$ fulfilling \eqref{wbs}  is a real pole of $G(z)$ (and viceversa). Thus the search for atom-photon BSs reduces to finding the real poles of $G(z)$. Assuming that $G_B(z)$ does not have real poles (\eg when $B$ is translationally invariant, see Section \ref{sec-TIB}), then we see from \eq\eqref{Green} that the poles of $G(z)$ are the real zeros of function $F(z)$ or equivalently the real roots of equation $F(\omega)=0$ (defined on the real $\omega$-axis). Using \eqref{fz} this equation more explicitly reads
	\begin{equation}
		\omega = \omega_0 + \bar{g}^2\braket{\chi| {G}_B ( \omega^+ ) |\chi}\,, 	\label{poleeq}
	\end{equation}
	where
	\begin{equation}
		\Sigma(\omega^+)=\bar{g}^2\braket{\chi| {G}_B ( \omega^+) |\chi} \label{self}
	\end{equation} 
	is naturally interpreted as the self-energy of the giant atom. The presence of $\omega^+$ (instead of $\omega$) is due in order to encompass BSs {\it in the continuum} (typically within a photonic band): as we will see later, these cannot occur with normal atoms but are possible with giant atoms (for BSs out of the continuum, $G_B(\omega^+)\equiv G_B(\omega)$ and subscript subscript "+" can be dropped). For simplicity, in the remainder we will refer to $\braket{\chi| {G}_B ( \omega^+) |\chi} $ (without the $\bar{g}^2$ factor) as the self-energy although it has of course the dimensions of the inverse of an energy. 
	
	In line with the general Green's function theory, an atom-photon BS (strictly speaking its associated projector $\ket{ \Psi_{\rm BS}} \!\bra{ \Psi_{\rm BS}}$) is the residue of $G(z)$ at a pole $z=\omega_{BS}$ fulfilling \eqref{poleeq}. Accordingly, with the help of \eqs \eqref{Green}, \eqref{psiz} and \eqref{psiiz} we see that a dressed BS for a giant atom has the general form
	\begin{equation}
		\ket{\Psi_{\rm BS}}=	\ket{\Psi(z{=}\omega_{\rm BS})}={\cal N} \left(\ket{e}+\ket{\psi_{\rm BS}}\right)\,\label{psibs1}
	\end{equation} 
	with the (unnormalized) single-photon state $\ket{\psi_{\rm BS}}$ and normalization factor ${\cal N}$ respectively given by
	\begin{eqnarray}
		\ket{\psi_{\rm BS}}&= \bar{g} \,G_B (\omega_{\rm BS}^+)\ket{\chi}\,, \label{psibs}\\
		{\cal N} &=\frac{1}{\sqrt{1+\braket{\psi_{\rm BS} |\psi_{\rm BS}}}}\,,\label{N-fac}
	\end{eqnarray} 
	and where (as said) $\omega_{\rm BS}$ is a solution of the pole equation \eqref{poleeq}. Notice that $\ket{\psi_{\rm BS}}$ is of the first order in the effective coupling strength ${\bar g}$ and, accordingly, ${\cal N}$ features no first order terms $\sim {\bar g}$.
	
	Replacing the expansion of $\ket{\chi}$ in terms of single-photon states $\{\ket{x}\}$ [see \eq\eqref{chi2}], we get the BS photonic wave function in real space
	\begin{equation}
		\ket{\psi_{\rm BS}}=\bar{g}\sum_{\ell=1}^{\cal N} \alpha_\ell \,G_B (\omega_{\rm BS}^+)\ket{x_\ell}\,.\label{psibs2}
	\end{equation} 
	For a normal atom placed at $x_\ell$, this reduces to $G_B (\omega_{\rm BS}^+)\ket{x_\ell}$ (unnormalized). Note that this might suggest that the giant-atom BS is a coherent superposition of normal-atom BSs, which is instead generally false since it would require $\omega_{\rm BS}$ to simultaneously fulfil \eqref{poleeq} and $\omega_{\rm BS}=\omega_{0}+\bar{g}^2 \braket{x_\ell| {G}_B ( \ombs ) |x_\ell}$ for any $\ell=1,...,{\cal N}$ which is not necessarily true.

	\subsection{In-gap bound states (BSs out of the continuum)}\label{sec-ig-bs}
	
	Inside a bandgap, $G_B(\omega^+)=G_B(\omega)$ [see \eqs\eqref{poleeq}] hence the giant-atom self-energy $\braket{\chi| {G}_B ( \omega ) |\chi}$ takes {\it real} values. 
	Using that ${G}_B ( \omega )=(\omega{-}H_B)^{-1}$, we then have $G_B'(\omega)=-(\omega-H_B)^{-2}$, where the prime denotes the derivative with respect to $\omega$. Thus
	\begin{equation}
		\frac{d}{{d}\omega}	\braket{\chi| {G}_B ( \omega ) |\chi} =	-\braket{\chi| (\omega-H_B)^{-2} |\chi} \le 0 \,.
	\end{equation}
	This implies that function $F(\omega)$ [\cf\eq\eqref{fz}] is {\it monotonic}. 
	Recalling now that the solutions of the pole equation \eqref{poleeq} are the zeros of $F(\omega)$, we conclude that {\it at most one BS per bandgap can exist}. 
	
	We point out that this property is independent of the number of coupling points. Therefore, at variance with in-band BSs (see next subsection), having a giant atom instead of a normal one does not affect the maximum number of BSs occurring in bandgap.

	\subsection{In-band bound states (BSs in the continuum)}\label{bic-section}
	For $\omega$ inside a continuous {\it band}, the self-energy $\braket{\chi| {G}_B ( \omega^+ ) |\chi}$ is now {\it complex} with real and imaginary parts given by 
	\begin{eqnarray}
		{\rm Re}\braket{\chi| {G}_B ( \omega^+) |\chi}&={\cal P}\int\! { \rm d } \omega' \frac{\rho ( \omega' )}{ \omega - \omega'} \,,\label{ReG}\\
		{\rm Im}\braket{\chi| {G}_B ( \omega^+) |\chi}&=-  \pi \rho ( \omega )\,\label{ImG}
	\end{eqnarray}
	with $\cal P$ the integral's principal value and
	\begin{equation}
		\rho ( \omega )=  \sum_{n,k} \delta ( \omega - \omega_{n{\bf k }} ) \,\braket{\chi | \phi_{n{\bf k}}}\braket{\phi_{n{\bf k}} | \chi  }\,, \label{DOSchi}
	\end{equation}
	where we used the property $1/y^+=1/(y+i\epsilon)={\cal P}(1/y) - i \pi \delta(y)$  with $\epsilon\rightarrow 0^+$ (see \ref{app0} for details). 
	The energy function \eqref{DOSchi} can be seen as the effective local density of states (LDOS), and coincides with the LDOS of a normal atom placed at the fictious position corresponding to the site state $\ket{\chi}$.
	Note that the sum over bands (index $n$) appears since in general there may be overlapping bands. 
	
	Since we look for {\it real} solutions $\omega_{ \rm BS}$ of the pole equation, these need to simultaneously satisfy the pair of equations
	\begin{eqnarray}
		&\omega_{ \rm BS} = \omega_0 + \bar{g}^2 \,{\rm Re} \braket{\chi| {G}_B ( \omega_{\rm BS}^+ ) |\chi}\,,\label{re-eq}\\
		&{\rm Im} \braket{\chi| {G}_B ( \omega_{\rm BS}^+) |\chi}=0\,\label{im-eq}
	\end{eqnarray}
	(in a bandgap the latter condition is guaranteed since the self-energy is real as shown in Section \ref{sec-ig-bs}). 
	
	This implies, in particular, that the energy of a BS must belong to the set of {\it zeros} of function ${\rm Im} \braket{\chi| {G}_B ( \omega^+) |\chi}$ (imaginary part of the self-energy). Now, it is clear from \eqref{ImG} that ${\rm Im} \braket{\chi| {G}_B ( \omega^+ ) |\chi}$ vanishes if and only if 
	\begin{equation}
		\braket{\chi | \phi_{n\bf k}} = 0\,\,\,{\rm for}\,\,\,\omega_{n\bfk}=\omega\label{ImGzero}\,.
	\end{equation}
	In other words, a {\it necessary} condition for having a BS of energy $\omega_{\rm BS}$ is that all the bath eigenstates of energy $\omega_{\rm BS}$ do not overlap the site state $\ket{\chi}$. More explicitly, this condition reads [see \eqs\eqref{chi2} and \eqref{phik1}]
	\begin{equation}
		\langle \chi\ket{\phi_{n{\mathbf k}}}=\sum_\ell \alpha^*_{\ell} \,u_{n\beta_{\ell}}(\mathbf k) \,e^{i {\mathbf k}\cdot \bfr_{\it \ell}}=0\,\,\,\label{cond-phik}
	\end{equation}
	with $|x_\ell\rangle$ ($\ell$th coupling point) here embodied by $\ket{ {\mathbf r}_\ell,\beta_\ell} $.
	Clearly, \eq\eqref{cond-phik} can never be satisfied by a normal atom since in this case \eqref{cond-phik} would feature only one term in the summation thus requiring a vanishing Bloch wave function, which is absurd.

	By plugging \eqref{chi2} into \eq\eqref{self} we get
	\begin{eqnarray}
		\braket{\chi| {G}_B ( \omega^+ ) |\chi}=&\sum_{\ell} |\alpha_\ell|^2\bra{x_\ell}{G}_B ( \omega^+ )\ket{x_\ell}\nonumber\\
		&+\sum_{\ell,\ell'} \alpha_{\ell}^*\alpha_{\ell'} \bra{x_\ell}{G}_B ( \omega^+ )\ket{x_{\ell'}}\,.\label{Goff}
	\end{eqnarray}
	This shows that the giant-atom self-energy depends in particular on {\it off-diagonal} matrix elements of the field's Green's function (terms $\ell{\neq} \ell'$).
	This marks a major difference from a normal atom (coupled to only one cavity), in which case only a single diagonal entry of $G_B(\omega)$ is involved. Such terms generally introduce an {\it oscillatory} dependence of the self-energy on energy $\omega$ (in contrast with a bandgap where it is monotonic), entailing that \eqs\eqref{re-eq} and \eqref{im-eq} can admit more than one solutions, meaning in particular that {\it multiple} BSs can occur inside a photonic band. This property, which was proven in detail in the specific case of a coupled-cavity array \cite{lim2023oscillating}, is key to the appearance of stationary oscillations exhibited by a giant atom with three coupling points in a waveguide \cite{KockumPRR20}.

	\subsection{Bound state in the weak-coupling regime }\label{sec-weak}
	
	When the strength $\bar g$ of the atom-field interaction is small enough and provided that $G_B(z)$ has a smooth behaviour around $z=\omega_{0}$ (meaning that we are in the weak-coupling regime), the existence conditions of a BS [\ie \eqs\eqref{re-eq} and \eqref{im-eq}] to leading order reduce to
	\begin{eqnarray}
		&\omega_{\rm BS} = \omega_0,\label{re-eq-a}\\
		&{\rm Im} \braket{\chi| {G}_B ( \omega_0^+) |\chi}=0\,.\label{im-eq-a}
	\end{eqnarray}
	Recalling \eqs \eqref{ImG} and \eqref{ImGzero}, the latter equation is equivalent to
	\begin{equation}
		\braket{\chi | \phi_{n\bf k}} = 0\,\,\,{\rm for}\,\,\,\omega_{n\bfk}=\omega_{\rm BS}=\omega_0\label{ImGzero2}\,,
	\end{equation}
	which thus embodies a necessary and sufficient condition for a BS to occur in the weak-coupling regime. Out of the continuum (\eg inside bandgaps), this is surely satisfied (due to lack of $H_B$'s eigenstates with energy $\omega_{0}$) and the weak-coupling BS always exists. In the continuum, instead, \eq\eqref{ImGzero2} cannot be satisfied by a normal atom 
	[recall \eq\eqref{cond-phik}] but can be matched by a giant atom with a suitable pattern of coupling points (\ie for a suitable site state $\ket{\chi}$). 
	
	When it exists, a BS in the weak-coupling regime has the form [\cf\eq\eqref{psibs1}]
	\begin{equation}
		\ket{\Psi_{\rm BS}}=  \ket{e}+ \ket{\psi_{\rm BS}}\,\,\,\,(\rm weak{-}coupling\,\,BS)\label{BS-weak}
	\end{equation}
	with
	\begin{eqnarray}
		\omega_{\rm BS}= \omega_{0}\,,\,\,\,\ket{\psi_{\rm BS}}= \bar{g} \,G_B (\omega_{0}^+)\ket{\chi}\,. \label{BS-weak2}
	\end{eqnarray} 
	Notice that $\ket{\Psi_{\rm BS}}$ is normalized to leading order since, as we observed earlier, \eqref{N-fac} features no first-order terms in $\bar{g}$. Also, note that $\ket{\Psi_{\rm BS}}\rightarrow \ket{e}$ in the limit $\bar{g}\rightarrow 0$.
	\\
	\\
	These BSs, which can show up even in the continuum with giant atoms \cite{gonzalez2019engineering,lim2023oscillating,soro2023interaction}, are crucial for the occurrence of decoherence-free interactions, as will become clear later in Section \ref{sec-DFH}.

	\section{Many giant atoms: Hamiltonian and master equation} \label{sec-ME}
	We now relax the assumption that only one giant atom is coupled to bath $B$ and consider now $N_a \ge 1$ giant atoms interacting with the field. Hamiltonian \eqref{H1} is naturally generalized as
	\begin{equation}
		H=H_B+\omega_0 \sum_{j=1}^{N_a}\sigma_{j+}\sigma_{j -}+\sum_{j=1}^{N_a}\sum_{\ell=1}^{\cal N} \left(g_{j \ell} b_{x_{j \ell}}^\dag \sigma_{j -}+\Hc\right)\label{H1-many},
	\end{equation}
	with $g_{j \ell}$ the coupling strength of the $\ell$th coupling point of the $j$th atom, this point having coordinate $x_{j \ell}$  (for simplicity we assume that the number of coupling points ${\cal N}$ and atom's frequency $\omega_{0}$ are the same for all atoms). 
	
	The site ladder operator \eqref{achi} becomes now atom-dependent 
	\begin{equation}
		b_{\chi_j}=\sum_{\ell=1}^{\cal N} \alpha_{j \ell}^* b_{x_{j \ell}}\,\,\,{\rm with}\,\,\alpha_{j \ell}=\frac{g_{j \ell}}{\bar{g}_j}\,\,{\rm and}\,\,\bar{g}_j=\sqrt{\sum_\ell |g_{j \ell}|^2}\,,\label{achi-2}
	\end{equation}
	where from now on we set $\bar{g}_j=\bar{g}$ (independent of the atom). Accordingly, the site state of atom $j$ now reads [\cf\eq\eqref{chi2}]
	\begin{equation}
		\ket{\chi_j} =b_{\chi_j}^\dag |0\rangle= \sum_{\ell=1}^{\cal N} \alpha_{j \ell} \ket{x_{j \ell}}\,.\label{chi2-2}
	\end{equation}
	When $B$ is a photonic lattice, this is written as
	\begin{equation}
		\ket{\chi_j} = \sum_{\ell=1}^{\cal N} \alpha_{j \ell} \ket{\mathbf{r}_{j \ell} ,\beta_{j \ell}}\,\label{chi2-3}
	\end{equation}
	with $\mathbf{r}_{j \ell}$ identifying the unit cell and $\beta_{j {\ell }}$ the sublattice of the $\ell$th coupling point of atom $j$.
	The total Hamiltonian can thus be written compactly as [\cf\eq\eqref{H1-bis}]
	\begin{equation}
		H=H_B+\omega_0 \sum_{j=1}^{ N_a} \sigma_{j+}\sigma_{j-}+\sum_{j=1}^{ N_a} \bar{g } \left(b_{\chi_j}^\dag \sigma_{j -}+\Hc\right)\label{H1-bis-2}
	\end{equation}
	(henceforth, it will be understood in all sums that index $j$ runs from 1 to $N_a$).
	
	Based on Hamiltonian \eqref{H1-bis-2}, at zero temperature and in the usual Markovian regime, the reduced state of the atoms $\rho$ at time $t$ obeys the Lindblad master equation (see \ref{appA})
	\begin{equation}
		\dot \rho=-i [H_{\rm eff},\rho]+ {\cal D}[\rho]\label{ME}
	\end{equation}
	with the effective Hamiltonian and collective dissipator given by
	\begin{eqnarray}
		H_{\rm eff}&=\sum_{j,j'}(\omega_{0} \delta_{jj'}+{\cal K}_{jj'})\sigma_{j+} \sigma_{j'-}\,,\label{Heff}\\
		{\cal D}[\rho]&=\sum_{j,j'} \gamma_{jj'}  \left[  \sigma_{j'-}  \rho \sigma_{j+} -\frac{1}{2}\, \left\{ \rho ,\sigma_{j+}  \sigma_{j'-} \right\} \right]\,, \label{Drho}
	\end{eqnarray}
	where
	\begin{eqnarray}
		{\cal K}_{j j'} & = \bar{g}^2\,\frac{\langle\chi_j |{G}_B ( \omega_0^+ )|\chi_{j'}\rangle+\langle\chi_{j'} |{G}_B ( \omega_0^+)|\chi_{j}\rangle^*}{2}\, \,,\label{Js}\\
		\gamma_{j j'}& = i \bar{g}^2\!\left(\braket{\chi_j |{G}_B ( \omega_0^+)|\chi_{j'}} {-}\braket{\chi_{j'} |{G}_B ( \omega_0^+ )|\chi_{j}}^* \right)\label{gam}.
	\end{eqnarray}
	
	More explicitly, rates ${\cal K}_{j j'}$ and $\gamma_{j j'}$ can be written as (see \ref{appA}) 
	\begin{eqnarray}
		{\cal K}_{jj'}&=\bar{g}^2\,{\cal P}\int\! { \rm d } { \omega}\,\frac{\rho_{j j'} ( \omega ) }{ \omega_0 - \omega } \label{jj1}\,\,,\\
		\gamma_{jj'} &=2 \pi \,\bar{g}^2  \rho_{j j'} ( \omega_0 )\,,\label{gg1}
	\end{eqnarray}
	where, similarly to  \eq\eqref{DOSchi}, we defined $\rho_{j j'} ( \omega ) = \sum_{n,\mathbf{k}} {\delta( \omega - \omega_{n\mathbf{k}} )} \langle \chi_j \ket{\phi_{n\mathbf{k}}} \!\bra{\phi_{n\mathbf{k}}} \chi_{j'} \rangle $, where we recall that   $\ket{\phi_{n{\bf k}}}$ are the eigenstates of $H_B$ and   $\omega_{n\bfk}$ their energies.

	\eq\eqref{ME} is a many-emitter master equation (holding even if bath $B$ is not translationally invariant), which is expressed in terms of the bath Green's function (master equations in this form are known for normal atoms, see \eg\rref \cite{asenjo2017exponential} considering a continuous photonic bath). 
	For normal atoms (${\cal N}=1$), the effective atom-atom coupling strengths ${\cal K}_{jj'}$ and dissipation rates $\gamma_{jj'}$ depend on two-point matrix elements of the bath Green's function involving the actual positions $x_j$'s of the atoms. \eq\eqref{Heff} shows that this remains {\it formally} true with giant atoms, but now in terms of their fictitious positions $\chi_j$'s [\cf definition \eqref{chi2-2}].
	
	We next focus on occurrence of decoherence-free effective Hamiltonians, namely those cases when all dissipation rates $\gamma_{j j'}$ in the above master equation vanish, which we will then link to atom-photon BSs.
	
	\section{Decoherence-free effective Hamiltonian}\label{sec-DFH}
	
	When the dissipator \eqref{Drho} vanishes, one is left with an effective Schr\"odinger equation for the giant atoms having as generator the effective many-body spin Hamiltonian $H_{\rm eff}$ [\cf\eq \eqref{Heff}] which is then referred to as a {\it decoherence-free effective Hamiltonian} (DFH). This occurs when rates $\gamma_{j j'}=0$ (for any $j$ and $j'$).

	As shown by \eq\eqref{gg1}, when $j'{=}j$ rate $\gamma_{j j'}$ clearly vanishes if and only if $\braket{ \chi_j | \phi_{n\bfk} } = 0$ for any $j$ and any $|\phi_{n{\bf k}} \rangle$ such that $\omega_{n\bfk}=\omega_{0}$, which then implies that $\gamma_{j j'}=0$ even for $j'\neq j$.  By recalling \eq\eqref{ImGzero2} (condition for occurrence of BS under weak coupling), we thus conclude that {\it a decoherence-free effective Hamiltonian arises if and only if each atom seeds the weak-coupling BS} [\cf\eqs\eqref{BS-weak} and \eqref{BS-weak2}]
	\begin{equation}
		\ket{\Psi^j_{\rm BS}}=  \ket{e_j}+ \ket{\psi^j_{\rm BS}}\,\,(\rm weak{-}coupling\,\,BS\,\,of\,\,atom\,\,{\it j})\label{BSmany}
	\end{equation}
	with $\ket{e_j}$ the state where the $j$th atom is excited and there are no photons. 
	To our knowledge, this property had not been highlighted in such an explicit way to date even for normal atoms. 
	
	Also, using \eqs \eqref{Heff}, \eqref{Js}, \eqref{gam} and \eqref{BSmany}, the decoherence-free effective Hamiltonian can be fully expressed in terms of the giant-atom BSs as
	\begin{equation}
		H_{ \rm eff} =  \sum_{j , j' } (\omega_0 \delta_{j j'} +  {\cal K}_{jj'} )\sigma_{j+} \sigma_{j'-}\,\,\,{\rm with}\,\,{\cal K}_{jj'}=\bar{g} \braket{ \chi_j | \psi_{\rm BS}^{j'}}\label{Heff2}\,,
	\end{equation}
	where we used that [\cf\eq\eqref{gam}] $\gamma_{j j'}=0$, entailing $\braket{\chi_{j'} |{G}_B ( \omega_0^+ )|\chi_{j}}^*=\braket{\chi_j |{G}_B ( \omega_0^+)|\chi_{j'}}$. Recall that $\ket{\psi_{\rm BS}^{j'}}\sim \bar{g}$, which shows explicitly that, as expected, ${\cal K}_{jj'}\sim \bar{g}^2$.
	
	It is worth noticing that $H_{ \rm eff}$ is guaranteed to be Hermitian since  ${\cal K}_{jj'}={\cal K}_{j'j}^*$ [which follows immediately from \eq\eqref{Js}]. This also entails (recall that $\bar{g}>0$)
	\begin{equation}
		\braket{ \chi_j | \psi_{\rm BS}^{j'}}=\braket{ \chi_{j'} | \psi_{\rm BS}^{j}}^*\,,\label{recipro}
	\end{equation}
	showing that the interaction strength ${\cal K}_{jj'}$ can be equivalently seen as either the overlap between the BS of emitter $j'$ and the site state of emitter $j$ or as the c.c. of the overlap between the BS of emitter $j$ and the site state of emitter $j'$. This equivalence can be useful, as we will show later on.
	
	This shows that two atoms get effectively coupled provided that the BS seeded by each overlaps with the site
	state of the other one, with the corresponding interaction strength essentially measured by the overlap.  More explicitly, 
	\begin{equation}
		\label{Heff3}
		{\cal K}_{jj'}=\bar{g}\sum_{\ell}\alpha^*_{j \ell} \bra{x_{j \ell}}\psi_{\rm BS}^{j'}\rangle = \bar{g}^2\sum_{\ell,\ell'}\alpha^*_{j \ell}\alpha_{j' \ell'} \bra{x_{j \ell} |G_B(\omega^+_{0})}x_{j'\ell'}\rangle\,. 
	\end{equation}
	%	\begin{eqnarray}
		%		\begin{split}\label{Heff3}
			%			{\cal K}_{jj'}&=\bar{g}\sum_{\ell}\alpha^*_{j \ell} \bra{x_{j \ell}}\psi_{\rm BS}^{j'}\rangle \\
			%			&=\bar{g}^2\sum_{\ell,\ell'}\alpha^*_{j \ell}\alpha_{j' \ell'} \bra{x_{j \ell} |G_B(\omega^+_{0})}x_{j'\ell'}\rangle\,. 
			%		\end{split}
		%	\end{eqnarray}
	To sum up, we have extended to giant atoms the property that decoherence-free interactions are always mediated by atom-photon BSs, either in or out of the continuum.
	
	We next proceed to the study of an important class of giant-atom BSs called "vacancy-like dressed states (VDSs)", the main motivation being that, combined with the general theory just shown, they provide a powerful tool to understand and predict giant-atom DFHs (as we show later on).

	\section{Vacancy-like dressed state (VDS)}\label{sec-vds}
	
	\subsection{Review of vacancy-like dressed states for normal atoms}\label{sec-review}
	
	For a normal atom coupled to a bath $B$, a VDS \cite{LeonfortePRL2021} is by definition a single-excitation dressed state 
	having the same energy as the atom, \ie
	\begin{equation}
		H \ket{\Psi_{\rm VDS}}=\omega_{0} \ket{\Psi_{\rm VDS}}\,,\label{vds-def}
	\end{equation}
	where $H$ is a special case of \eqref{HB} for ${\cal N}=1$ and $g_\ell=g$.
	Notice that, by definition, identity \eqref{vds-def} exactly holds for {\it any} value of the coupling strength $g$ (not only in the weak-coupling regime as in Section \ref{sec-weak}), meaning in particular that even if $g$ is made larger and larger the VDS energy remains pinned to the bare atom energy $\omega_0$. VDSs enter basic waveguide-QED phenomena such as perfect reflection of a photon from an atom and dressed BSs in the continuum \cite{LeonfortePRL2021} (recently, they were extended to fermionic matter-wave systems \cite{windt2023fermionic}). Moreover, they are essential to understand occurrence of atom-photon BSs enjoying topological/symmetry protection \cite{Bello2019,LeonfortePRL2021,VegaPRA21,bello2}.
	\begin{figure}[t!]
		\begin{center}
			\includegraphics[width=0.75\textwidth]{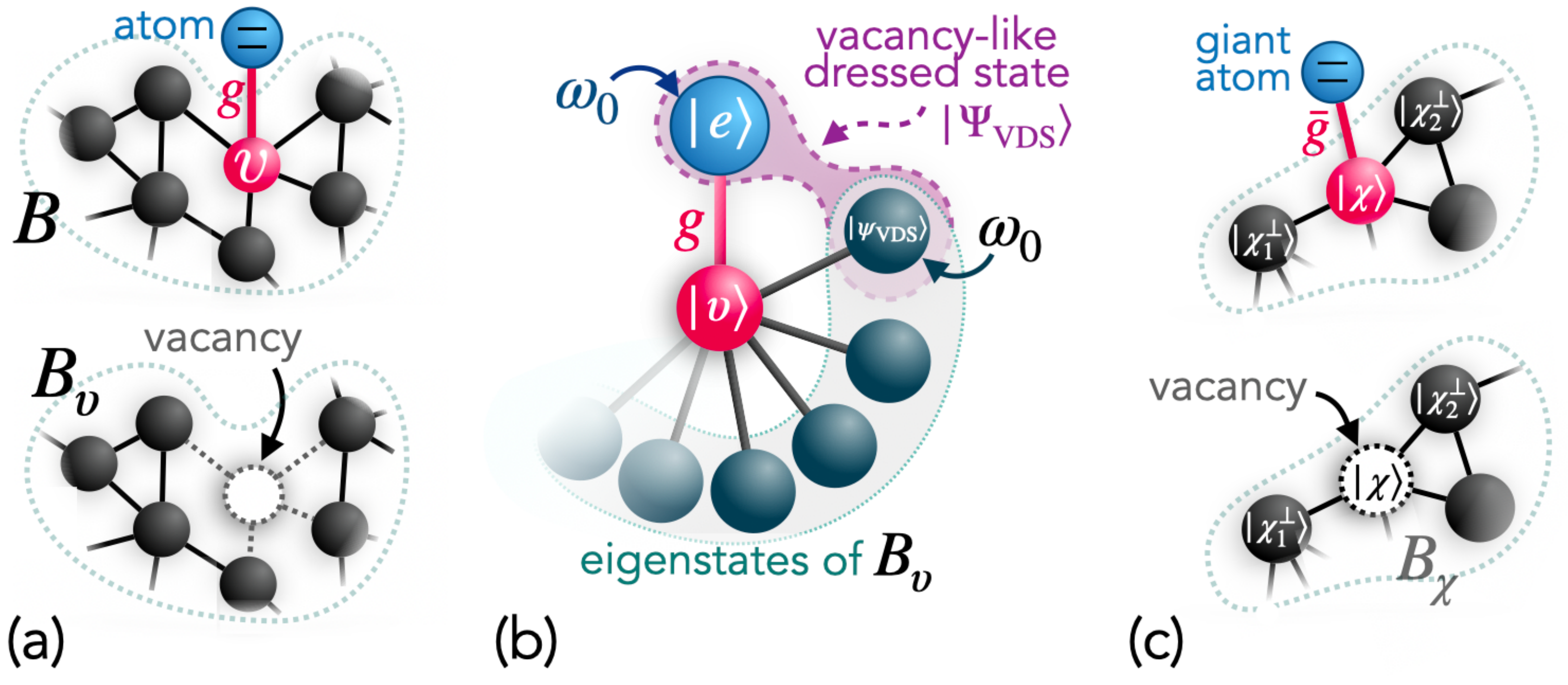}
			\caption{Vacancy-like dressed states (VDSs). (a) A normal atom is coupled to cavity $v$ of an unspecified bath $B$ [whose Hamiltonian $H_B$ is given by \eq\eqref{HB}]. The bath where cavity $v$ is replaced by a vacancy is called $B_v$ and its free Hamiltonian $H_{B_v}$. (b) Formation mechanism of a VDS: if there exists an $H_{B_v}$'s eigenstate called $\ket{\psi_{\rm VDS}}$ with energy $\omega_{0}$, there always exists a superposition of $\ket{e}$ and $\ket{\psi_{\rm VDS}}$ that decouples from $\ket{v}$. This superposition, called $\ket{\Psi_{\rm VDS}}$, is thus an eigenstate of the total Hamiltonian $H$ with eigenvalue $\omega_0$. (c) VDSs are naturally extended to a giant atom, where the roles of $\ket{v}$ and $B_v$ are now respectively played by the site state $\ket{\chi}$ and $B_\chi$ (the Hilbert space of $B_\chi$ is the set of all single-photon states which are orthogonal to $\ket{\chi}$).\label{fig-VDS}} 
		\end{center}
	\end{figure}
	
	While such states can be both bound and unbound, in this paper we focus solely on {\it bound} VDSs, hence from now on "VDS" must be always intended as a bound VDS (\ie $\langle \Psi_{\rm VDS}\ket{\Psi_{\rm VDS}}=1$).
	
	Notably, there exists a tight relationship between VDSs and bound states induced by a vacancy, hence the name “vacancy-like dressed states”. To see this, in line with \rref\cite{LeonfortePRL2021}, we call "$v$" the cavity directly coupled to the normal atom and $B_v$ the bath $B$ with a vacancy in place of cavity $v$ [see \fig \ref{fig-VDS}(a)]. Then it can be shown that any VDS can be expressed in the form
	\begin{equation}
		\ket{\Psi_{\rm VDS}}=\cos\theta\,|e\rangle+e^{i \varphi} \sin\theta \ket{\psi_{\rm VDS}}\,,\label{dressed}
	\end{equation}
	with
	\begin{eqnarray}
		\theta=\arctan{|\eta|}\,,\,\,\varphi=\arg{\eta}\,\,\,\,\,{\rm with}\,\,\eta=-\frac{g}{\langle v |H_B|\psi_{\rm VDS}\rangle}\,\label{angle}
	\end{eqnarray}
	and where, importantly, $\ket{\psi_{\rm VDS}}$ is a bound state seeded by a vacancy at site $v$, \ie a normalized eigenstate of  $H_{B_v}$ (free Hamiltonian of $B_v$) 
	\begin{equation}
		H_{B_v}\ket{\psi_{\rm VDS}}=\omega_{0}\ket{\psi_{\rm VDS}}\,.\label{Hbv}
	\end{equation}
	Notice that the eigenvalue of $\ket{\psi_{\rm VDS}}$ here is the same as the one in \eqref{vds-def} and coincides with the atom's frequency.
	To understand occurrence of a VDS, notice that $\ket{v}$ is coupled to both $\ket{e}$ and bath $B_v$ as shown in \fig \ref{fig-VDS}(b). By decomposing $H_{B_v}$ into its eigenstates [see \fig \ref{fig-VDS}(b)], one obtains a star-like configuration having $\ket{v}$ at its center. Now, through a mechanism in fact analogous to formation of dark states \cite{LP07}, if one of the $H_B$'s eigenstates has energy $\omega_{0}$ (same as $\ket{e}$) then there always exists a superposition of this eigenstate (called $\ket{\psi_{\rm VDS}}$) and $\ket{e}$ which decouples from $\ket{v}$ and thereby is a stationary state of the total system with energy $\omega_{0}$ (dressed state) \cite{LeonfortePRL2021}. 
	\\
	\\
	Reversing the above picture provides a practical method to find a VDS as follows: one first searches for a (bound) $H_{B_v}$'s eigenstate $\ket{\psi_{\rm VDS}}$ such that $\langle v |H_B|\psi_{\rm VDS}\rangle\neq 0$ and next tunes the atom to the energy of this $H_B$'s eigenstate. The superposition of $\ket{\psi_{\rm VDS}}$ and $\ket{e}$ defined by \eqref{dressed} is then a VDS.

	\subsection{Giant-atom VDS}\label{VDS-sec}
	
	A giant-atom VDS is naturally defined formally just as in \eq\eqref{vds-def}. Since in the picture defined by transformation $U_\chi$ (see Section \ref{sec-uchi}) the giant atom is effectively turned into a normal atom coupled only to $\ket{\chi}$, we can formally apply the standard VDS theory just reviewed but with $g$ and $\ket{v}$ now replaced by $\bar{g}$ and $\ket{\chi}$,  respectively [\cf\eqs\eqref{achi} and \eqref{chi2}]. 
	Accordingly, $B_v$ (bath with a vacancy substituting $v$) [see \fig\ref{fig-VDS}(a)] is now replaced by $B_\chi$ [see \fig\ref{fig-VDS}(c)] namely the set of all single-photon states which are othogonal to the site state $\ket{\chi}$.
	
	Thereby, a VDS has the general form \eqref{dressed}, where now
	\begin{eqnarray}
		\theta=\arctan{|\eta|}\,,\,\,\varphi=\arg{\eta}\,\,\,\,\,{\rm with}\,\,\eta=-\frac{\bar g}{\langle \chi |H_B|\psi_{\rm VDS}\rangle}\,\label{angle-chi}
	\end{eqnarray}
	and where, importantly, $\ket{\psi_{\rm VDS}}$ now is a normalized eigenstate of $H_{B_\chi}$ (free Hamiltonian of $B_\chi$) with energy $\omega_0$
	\begin{equation}
		H_{B_\chi}\!\ket{\psi_{\rm VDS}}=\omega_{0}\ket{\psi_{\rm VDS}}\,.\label{Hbchi}
	\end{equation}
	With the described replacements [see \fig\ref{fig-VDS}(c)], the formation mechanism of a giant-atom VDS is a natural extension of the one for a normal atom [\cf \fig\ref{fig-VDS}(b)].
	We stress that, like its normal-atom counterpart, a distinctive feature of a giant-atom VDS is that it arises -- always at frequency $\omega_0$ -- no matter how large the coupling strength $g$ (hence even in regimes when the atom decay is non-Markovian). 
	\\
	\\
	For future use (when we will discuss DFHs), it is useful to give here the expression of the VDS \eqref{dressed} in the weak-coupling limit, which reads 
	\begin{equation}
		\ket{\Psi_{\rm VDS}}=|e\rangle+\eta \ket{\psi_{\rm VDS}}
	\end{equation}
	with [\cf\eq\eqref{angle}] $\eta=-{\bar g}/{\langle \chi |H_B|\psi_{\rm VDS}\rangle}$. Accordingly, we can state that when a BS is also a VDS then \eqref{BS-weak} holds with [\cf\eq\eqref{BS-weak2}]
	\begin{equation}
		\ket{\psi_{\rm BS}}=\frac{\bar g}{\langle \chi |H_B|\psi_{\rm VDS}\rangle} \ket{\psi_{\rm VDS}}\,,\label{BS-weak2-vds} 
	\end{equation}
	hence (up to a constant factor) $\ket{\psi_{\rm BS}}$ essentially coincides with $\ket{\psi_{\rm VDS}}$\footnote{In the general case (beyond weak coupling), \eq\eqref{BS-weak2-vds} features an additional factor $1/{\cal N}$ on the right-hand side [\cf\eqs\eqref{psibs1} and \eqref{N-fac}].}. 
	It is worth to point out that for the above weak-coupling limit to exist we must require the natural condition $\langle \chi |H_B|\psi_{\rm VDS}\rangle\neq 0$ [otherwise $\eta\rightarrow \infty$, see \eq\eqref{angle-chi}].
	\\
	\\
	We already made clear in Section \ref{sec-DFH} that occurrence of DFHs goes hand in hand with formation of BSs. Searching for BSs can thus be exploited as a method to engineer novel types of DFHs with giant atoms. This is in particular true for VDSs, whose emergence in many systems can be predicted in a relatively straightforward fashion (as we will see through several examples in the next section) according to the following recipe. First, based on the pattern of coupling points, identifies the site state $\ket{\chi}$ [see \fig\ref{fig-setup}(b)]. All single-photon states orthogonal to $\ket{\chi}$ then define $B_\chi$ [see \fig\ref{fig-VDS}(c)]. Next, inside $B_\chi$ so identified, search for an $H_{B_\chi}$'s eigenstate $\ket{\psi_{\rm VDS}}$ such that $\langle \chi |H_B|\psi_{\rm VDS}\rangle\neq 0$.  The superposition of $\ket{\psi_{\rm VDS}}$ and $\ket{e}$ defined by \eqref{dressed} is then ensured to be a VDS, whose photonic wave function is given by \eqref{BS-weak2-vds}. 
	
	\subsection{Decay dynamics in the presence of a VDS}
	
	When a VDS exists, an initially excited quantum emitter will just not decay in the weak-coupling regime (in which case the atomic component of the VDS dominates over the photonic one). This will happen even if $\omega_0$ lies in the photonic continuum, in which case $\ket{e}$ is a {\it subradiant} state. 
	
	Beyond weak coupling, the VDS is still guaranteed to exist (see discussion in Section \ref{VDS-sec}) but will now feature a larger photonic fraction. Then, if no additional BSs arise, the atom will generally behave in a non-Markovian fashion exhibiting fractional decay, meaning even at long times the emitter keeps a residual excitation measured by $|\langle e|\Psi_{\rm VDS}\rangle|$.

	\section{Examples of giant-atom VDS}\label{sec-ex}
	
	In the following, we provide some paradigmatic examples of bound VDSs occurring with giant atoms.

	\subsection{Photonic graphene, three coupling points}\label{sec-gra1}

\begin{figure*}[t!]
		\begin{center}
			\includegraphics[width=1.\textwidth]{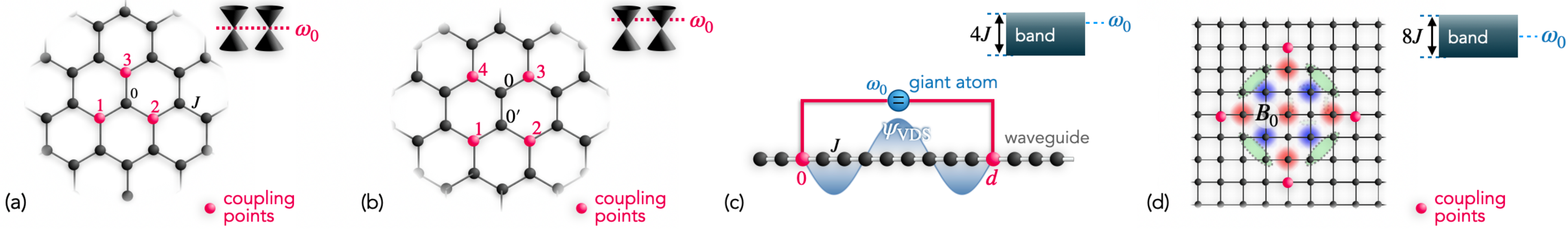}
			\caption{Examples of VDSs with giant atoms. (a) Photonic graphene and giant atom with three coupling points. (b) Same as panel (a) in the case of four coupling points. (c) Coupled-cavity array (discrete waveguide) and giant atom with two coupling points. (d) Homogeneous square lattice and giant atom with four coupling points. The VDS has non-zero amplitude only on red and blue cavities on which it takes uniform modulus while the phase is 0 ($\pi$) on red (blue) cavities. Each panel (top right) features also a sketch of the energy spectrum of $H_B$ and the required atom frequency $\omega_0$ for the considered VDS to occur.} \label{fig-vds-ex}
		\end{center}
	\end{figure*}

	Assume that $B$ is a 2D honeycomb lattice ("photonic graphene") [see \fig\ref{fig-vds-ex}(a)] with nearest-neighbour hopping rates $J$ and where each bare cavity frequency is $\omega_{c}$ [thus in \eqs\eqref{HB} and \eqref{HB2}, $\omega_x=\omega_c$ while $J_{xx'}=J$ for $x$ and $x'$ nearest neighbours while $J_{xx'}=0$ otherwise]. As shown \fig\ref{fig-vds-ex}(a), the bath is
	coupled to a giant atom with three coupling points (${\cal N}=3$) of equal strengths $g$ (thus $g_\ell=g$ for $\ell=1,2,3$), which are the three nearest neighbours of a certain cavity (no matter which). For convenience, we label the three coupling points simply with $\ell=1,2,3$ and the central cavity with 0 (see figure). Thus the site state in this case is the symmetric superposition of $\ket{1}$, $\ket{2}$ and $\ket{3}$ [\cf\eq\eqref{chi2}], \ie
	\begin{equation}
		\ket{\chi}=\frac{1}{\sqrt{3}}(\ket{1}+\ket{2}+\ket{3})\,.
	\end{equation}
	Clearly, state $\ket{0}$ lies fully within $B_\chi$ since it is orthogonal to $\ket{\chi}$. Moreover, $\ket{0}$ is coupled by $H_B$ to $\ket{\chi}$ since $\langle \chi|H_B|0\rangle=\sqrt{3} J$. Given that besides $\ket{\chi}$ there is no other single-photon state coupled to $\ket{0}$, it is evident that $\ket{0}$ is an eigenstate of $H_{B_\chi}$
	\begin{equation}
		H_{B_\chi} \!\ket{0}=\omega_{c}\ket{0}\,.
	\end{equation}
	Based on the last section, thereby, by setting $\omega_{0}=\omega_{c}$ a VDS \eqref{dressed} exists with $\ket{\psi_{\rm VDS}}=\ket{0}$ and [\cf\eq\eqref{BS-weak2-vds}]
	\begin{equation}
		\ket{\psi_{\rm BS}}=\frac{g}{J}\ket{0} \,,\label{VDS1} 
	\end{equation}
	where we used that $\bar {g}=\sqrt{3} g$ [\cf\eq\eqref{achi}].
	This VDS-BS thus features a photonic wave function fully localized in the region (a single cavity in the present case) surrounded by the three coupling points.
	
	We notice that $H_B$ has the well-known graphene energy spectrum [see top-right panel of \fig\ref{fig-vds-ex}(a)], featuring two bands that "touch" one another just at energy $\omega=\omega_{c}$ (corresponding to the well-known Dirac points). The above BS thus has a hybrid nature between a BS within a bandgap and a BS in the continuum.
	We notice that, at this energy, no BS is possible with a normal atom (local coupling), but only a "quasi-bound" state extending over a large region \cite{gonzalez2018exotic}.

	We next consider a giant atom with four coupling points $\ell=1,2,3,4$ (equal strengths) again coupled to the honeycomb lattice as shown in \fig\ref{fig-vds-ex}(b). The site state reads $\ket{\chi}=\frac{1}{2}(\ket{1}+\ket{2}+\ket{3}+\ket{4})$ and there are now {\it two} internal cavities called 0 and $0'$ which lie fully within $B_\chi$. Notice that, in addition to $\ket{\chi}$, state $\ket{0}$ ($\ket{0'}$) is now coupled by $H_B$ to $\ket{0'}$ ($\ket{0}$) and, most notably, also to states $\{\ket{\chi^{\perp}_i}\}$ (see Section \ref{sec-uchi}), \ie superpositions of $\{\ket{1},\ket{2},\ket{3},\ket{4}\}$ different from $\ket{\chi}$ (this is at variance with the previous case in Section \ref{sec-gra1} where the central site was coupled only to $\ket{\chi}$). Accordingly,
	\begin{equation}
		H_{B_\chi}\ket{0}= J \ket{0'}+\sum_i \!\bra{\chi^{\perp}_i}H_B|0\rangle\ket{\chi^{\perp}_i}\,,
	\end{equation}
	and an analogous equation holds by swapping $0$ with $0'$.
	It turns out that $\ket{\psi_+}=\frac{1}{\sqrt{2}}(\ket{0}+  \ket{0'})$ (symmetric superposition of $\ket{0}$ and $\ket{0'}$) fulfils 
	\begin{equation}
		H_{B_\chi } \!\ket{\psi_+}=(\omega_{c}+J)\ket{\psi_+}\,.
	\end{equation}
	This is because, due to reflection symmetry [see \fig\ref{fig-vds-ex}(b)], $\ket{\psi_+}$ is uncoupled from the three states $\ket{\chi^{\perp}_i}\}$, \ie $\bra{\chi^{\perp}_i}H_B|\psi_+\rangle=0$, while it is coupled to the site state $\ket{\chi}$ according to
	\begin{equation}
		\bra{\chi}H_B|\psi_+\rangle=\sqrt{2}J\,\,.
	\end{equation}
	According to the last section, since $\ket{\psi_+}$ is an eigenstate of $H_{B_\chi}$, by setting $\omega_{0}=\omega_{c}+ J$ a VDS \eqref{dressed} exists with $\ket{\psi_{\rm VDS}}=\ket{\psi_+}$ and [\cf\eq\eqref{BS-weak2-vds}]
	\begin{equation}
		\ket{\psi_{\rm BS}}=\frac{\sqrt{2}g}{J}\ket{\psi_+} \,, \label{psip}
	\end{equation}
	where used that $\bar {g}=2 g$ [\cf\eq\eqref{achi}].
	
	\subsection{Discrete waveguide, two coupling points}\label{sec-wav}
	
	Consider now a homogeneous coupled-cavity array (discrete waveguide) described by the Hamiltonian $H_B=-J \sum_ n |n\rangle\!\langle n+1|+{\rm H.c.}$ (the frequency of each cavity is set to zero), whose well-known energy spectrum features a single band of width $4J$ centered at $\omega=0$ [see inset of \fig\ref{fig-vds-ex}(c)].
	A giant atom with ${\cal N}=2$ is coupled with the same strength $g$ to the two cavities $n{=}0$ and $n{=}d$ [see \fig\ref{fig-vds-ex}(c)]. Thus (see Section \ref{sec-uchi})
	\begin{equation}
		\ket{\chi}=\frac{1}{\sqrt{2}}(\ket{0}+  \ket{d})\,,\,\,\,\ket{\chi^\perp}=\frac{1}{\sqrt{2}}(\ket{0}-  \ket{d})\,\,.\label{chi-wav}
	\end{equation} 
	Sites $n=1,...,d-1$, lying between the giant's atom's coupling points, can be jointly viewed as a coupled-cavity array of {\it finite} length $d$ (thus subject to standard open boundary conditions). Their corresponding free Hamiltonian is thus that of a discrete 1D cavity, whose well-known eigenstates and energy spectrum respectively read (see \eg \rrefs \cite{Longhi,LeonfortePRL2021})
	\begin{eqnarray}
		\ket{\psi_{k_m}}&=\sqrt{\frac{2}{d}}\,\,\sum_{n =1}^{d-1}\,\!\sin (k_m n) \ket{n}\,\label{modes-2-1}\,,\\
		\omega_{k_m}&=\omega_{c}-2J \cos k_m\label{omegak}
	\end{eqnarray}
	with $k_m=m\pi /d$ and $m=1,2,...,d-1$. By construction, states \eqref{modes-2-1} belong to $B_{\chi}$. We next note that
	$H_B$ enjoys reflection symmetry around the midpoint between sites 0 and $d$ and, moreover, that (with respect to the same symmetry) $\ket{\psi_{k_m}}$ has parity $(-1)^{m+1}$, while $\ket{\chi}$ ($\ket{\chi^\perp}$) has parity $\pm1$ [see \fig\ref{fig-vds-ex}(c)]. Thus, for $m$ {\it odd}  $|\psi_{k_m}\rangle$ has even parity and decouples from $\ket{\chi^\perp}$, \ie $\bra{\chi^\perp}H_B|\psi_{k_m}\rangle=0$. Hence,
	\begin{eqnarray}
		H_{B_\chi } \!\ket{\psi_{k_m}}=\omega_{k_m}\ket{\psi_{k_m}}\,\,\,{\rm for}\,\,m\,\,{\rm odd}\,\,.
	\end{eqnarray}
	
	Therefore, the condition to have a VDS \eqref{dressed} with $\ket{\psi_{\rm VDS}}=|\psi_{k_m}\rangle$ reads $\omega_{0}=\omega_{k_m}$ for some odd $m$. By introducing $k_0$ such that $\omega_{k_m=k_0}=\omega_{0}$, this condition reads $k_m=k_0$, namely (we set $m=2  \nu+1$ wth $ \nu$ integer since $m$ must be odd)
	\begin{equation}
		k_0 d=(2 \nu+1) \pi\,\,.\label{k0d}
	\end{equation}
	The BS photonic wave function then reads [\cf\eq\eqref{BS-weak2-vds}]
	\begin{equation}
		\ket{\psi_{\rm BS}}=-\sqrt{\frac{d}{2}}\,\frac{g}{\sin(k_m)J}\,\ket{\psi_{k_m}} \,, \label{psikm}
	\end{equation}
	where we used that $\bar {g}{=}\sqrt{2} g$ and $\bra{\chi}H_B|\psi_{k_m}\rangle{=}-\frac{2}{\sqrt{d}}\sin(k_m )J$.
	
	It can be shown that, under a suitable mapping ("waveguide unfolding") \cite{WitthautNJP10,FangNJP18}, the above BS is equivalent to the known BS in the continuum seeded by a normal atom in front of a mirror \cite{Longhi,TufarelliPRA13,TufarelliPRA14,Calajo2019}.

	\subsection{Square lattice, four coupling points}
	
	The most natural 2D extension of the previous coupled-cavity array is the square lattice in \fig\ref{fig-vds-ex}(d), where $J$ again denotes the nearest-neighbour hopping rate and we set to zero the frequency of each cavity ($\omega_{c}=0$). The energy spectrum is known to consist of a single band of width $8J$ centered at $\omega_c=0$ (where a singularity in the photonic density of states is known to occur). 
	
	We consider a giant atom with four coupling points $\ell=1,2,3,4$ (equal strength $g_\ell=g$) placed on the four vertices of a square \footnote{We consider a square for simplicity, but the present discussion can be naturally generalized to a rectangle.} like the one sketched in \fig\ref{fig-vds-ex}(d). Let $B_0$ be the set of cavities [nine overall in \fig\ref{fig-vds-ex}(d)] which lie inside the square internal to the coupling points. These cavities embody themselves a finite-size lattice called $B_0$ in the remainder, which evidently belongs to $B_\chi$. Unlike all previous instances, each $B_0$'s cavity is now generally coupled not only to $\ket{\chi}=\frac{1}{2}\sum_{\ell=1}^4 \ket{\ell}$ and $\{\ket{\chi_i^\perp}\}$ but even to the single-photon states corresponding to the boundary cavities inside the dashed ellipses in \fig\ref{fig-vds-ex}(d) (a total of eight sites in this example). 
	
	It is easy to see that the finite lattice $B_0$ admits a zero-energy eigenstate $\ket{\psi_0}$ whose wave function has non-zero amplitude only on red and blue cavities in \fig\ref{fig-vds-ex}(d), on which it takes uniform modulus and phase 0 ($\pi$) on red (blue) cavities. This state is explicitly written as \cite{windt2023fermionic1,Gonzalez-Tudela2017b}
	\begin{equation}
		\ket{\psi_{0}}=\frac{1}{\sqrt{N_0}}\left(\sum_{x\in P} \ket{x}-\sum_{x\in M} \ket{x}\right)\label{psi0}
	\end{equation}
	with $P$ ($M$) denoting the set of cavities where the phase is 0 ($\pi$) and $N_0=\mu^2$ (with $\mu$ an integer) the overall number of cavities where $\ket{\psi_{0}}$ has non-zero amplitude [\ie the total number of elements of the set $P\cup M$]. The example in \fig\ref{fig-vds-ex}(d) features $N_0=9$.

	It can be checked by inspection [see \fig\ref{fig-vds-ex}(d)] that, due to destructive interference, state $\ket{\psi_0}$ is effectively decoupled from the aforementioned "boundary" cavities and is only coupled to state $\ket{\chi}$ according to $\langle \chi|H_B| \psi_{0}\rangle=2J/\sqrt{N_0}$. Accordingly, setting $\omega_{0}=\omega_{c}=0$, a VDS \eqref{dressed} exists with $\ket{\psi_{\rm VDS}}=\ket{\psi_0}$ whose photonic wave function reads [\cf\eq\eqref{BS-weak2-vds}]
	\begin{equation}
		\ket{\psi_{\rm BS}}=\frac{g \sqrt{N_0}}{J} \ket{\psi_{0}}\label{psi02}\,.
	\end{equation}
	These results can be contrasted with a normal atom, which under analogous conditions ($\omega_0=\omega_c=0$) is not only unable to seed a BS but even shows up intrinsically non-Markovian emission due to the aforementioned singularity \cite{Gonzalez-Tudela2017b}. We note that the photonic wave function of the present giant-atom VDS is essentially analogous to the multi-atom bound state first investigated in \rrefs\cite{Gonzalez-Tudela2017b,Gonzalez-Tudela2017a} and recently extended to a fermionic bath with four impurities \cite{windt2023fermionic}.

	\section{VDS-mediated decoherence-free Hamiltonians}\label{sec-VDS-DFH}
	
	Based on Section \ref{sec-DFH}, each type of (bound) VDS derived in the previous section gives rise to a corresponding DFH (when the condition to seed such BS are met by each giant emitter). 
	
	\subsection{Photonic graphene}
	
\begin{figure*}[t!]
		\begin{center}
			\includegraphics[width=1\textwidth]{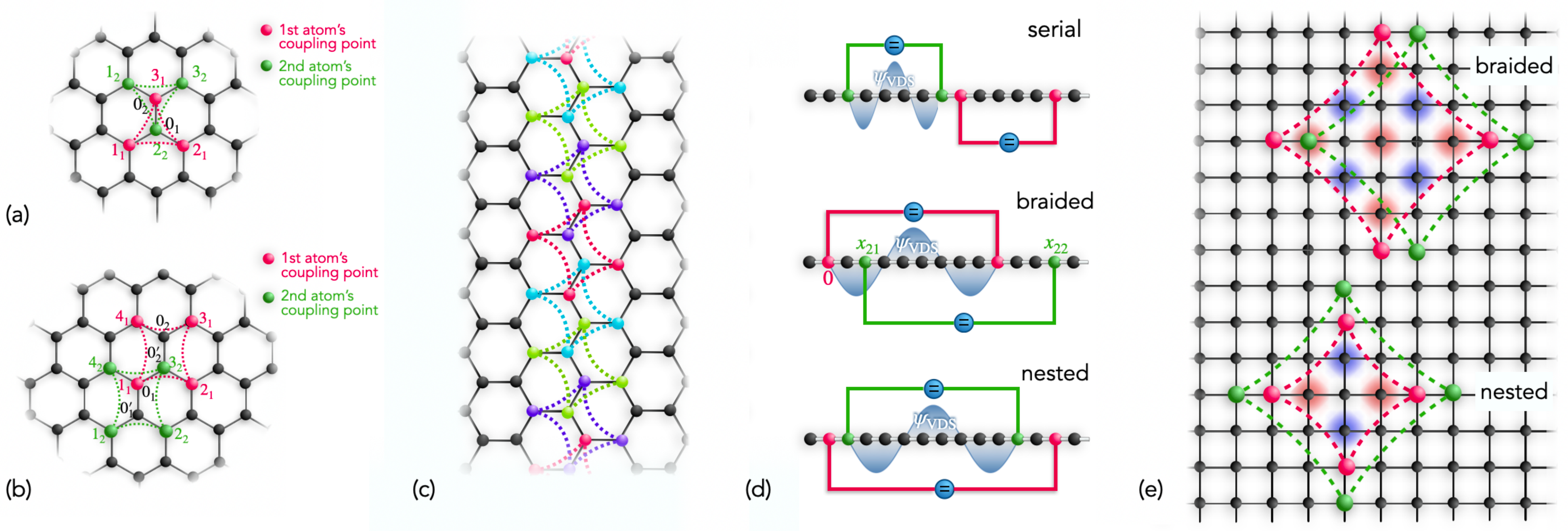}
			\caption{Examples of VDS-based DFHs with giant atoms. (a) Two giant atoms with three coupling points in photonic graphene [\cf\fig\ref{fig-vds-ex}(a)]. (b) Two giant atoms with four coupling points in photonic graphene [\cf\fig\ref{fig-vds-ex}(b)]. (c) Generalization of (b) to a periodic arrangment of giant atoms. (d) Pair of giant atoms with two coupling points in a discrete waveguide [\cf\fig\ref{fig-vds-ex}(c)] in the serial, braided and nested configurations. (e) Pair of giant atoms with four coupling points in a square lattice: braided and nested arrangements.} \label{fig-DFH-ex} 
		\end{center}
	\end{figure*}

	Assume to have a pair of identical giant atoms (labeled by $j=1,2$) with ${\cal N}=3$ and $\omega_0=\omega_{c}$, each coupled to photonic graphene according to the scheme shown in \fig\ref{fig-DFH-ex}(a): all the coupling points of emitter $j$ (here called $\ell_j$ with $\ell=1,2,3$ and $j=1,2$) are the nearest neighbours of one coupling point of emitter $j\neq j'$. The coupling strength of any copling point is $g$. We know that a DFH occurs if each atom in the weak-coupling regime gives rise to a BS with $\omega_{ \rm BS}\simeq \omega_{0}$ [\cf \eq\eqref{BS-weak} and Section \ref{sec-DFH}]. As shown in Section \ref{sec-gra1} and \fig\ref{fig-vds-ex}(a), this indeed happens when $\omega_0=\omega_{c}$ in which case each emitter $j=1,2$ seeds the VDS in Section \ref{sec-gra1} (so that $\omega_{\rm BS}=\omega_{0}$ is matched exactly) 
	with [\cf\eq\eqref{VDS1}]
	\begin{equation}
		\ket{\psi^j_{\rm BS}}=\frac{g}{J}\ket{0_j} \,\label{VDS1-2 }.
	\end{equation}
	Clearly, we have [see \fig\ref{fig-DFH-ex}(a)] $\ket{0_1}\equiv \ket{1_2}$ while $\ket{0_2}\equiv \ket{2_1}$. Hence, recalling that $\ket{\chi_j}=\frac{1}{\sqrt{3}}\sum_{\ell=1}^3 \ket{\ell_j}$ and ${\bar g}=\sqrt{3}g$, we get a DFH defined by \eq\eqref{Heff2} with
	\begin{equation}
		{\cal K}_{12}={\cal K}_{21}={\bar g}\,\langle \chi_1 | \psi_{\rm BS}^2\rangle=\frac{g^2}{J}\,\,.
	\end{equation}
	
	Similar conclusions hold if ${\cal N}=4$ and $\omega_0=\omega_{c}+J$ with the two giant atoms arranged as in \fig\ref{fig-DFH-ex}(b). Now, an argument analogous to the previous ${\cal N}=3$ case combined with the results of Section \ref{sec-gra1} and \fig\ref{fig-vds-ex}(b) yield [\cf\eq\eqref{psip}] $\ket{\psi^j_{\rm BS}}=\frac{\sqrt{2}g}{J}\ket{\psi_+^j}$ and hence
	\begin{equation}
		{\cal K}_{12}={\cal K}_{21}={\bar g}\,\langle \chi_1 | \psi_{\rm BS}^2\rangle=2\sqrt{2}\,\frac{g^2}{J}\langle \chi_1 | \psi_+^2\rangle=\frac{g^2}{J}\,\,.\label{j12}
	\end{equation}
	(we used $\ket{\chi_j}=\frac{1}{2}\sum_{\ell=1}^4\ket{\ell_j}$ and ${\bar g}=4 g$).
	\\
	\\
	The above schemes can be naturally scaled to many atoms. An instance is shown in \fig\ref{fig-DFH-ex}(c), which extends the scheme in panel (b) realizing an effective {\it one-dimensional} spin Hamiltonian with strictly {\it nearest-neighbour} interactions of strength equal to \eqref{j12}. Although not shown in the figure, it is evident that a {\it two-dimensional} spin Hamiltonian can be constructed by a natural 2D generalization of \fig\ref{fig-DFH-ex}(c).

	\subsection{Discrete waveguide}\label{dis-sec}
	
	Consider a pair of giant atoms with ${\cal N}=2$ each coupled to a discrete waveguide as in \fig\ref{fig-DFH-ex}(d) [compare with \fig\ref{fig-vds-ex}(c)], where cavities $0$ and $d$ are the coupling points of atom 1, while $x_{21}$ and $x_{22}$ generically denote the coupling points of atom 2 (three different choices are skecthed). According to Section \ref{sec-wav} [see also \eq\eqref{k0d}], under the conditions 
	\begin{eqnarray}
		k_0 d=(2 \nu+1) \pi\,,\,\,k_0 (x_{22}{-}x_{21})=(2 \nu'+1) \pi\,,\label{nup}
	\end{eqnarray}
	each emitter seeds a VDS. In particular, atom 1 seeds a VDS with photonic wave function [\cf\eq\eqref{psikm}]
	\begin{equation}
		\ket{\psi_{\rm BS}^1}=-\sqrt{\frac{d}{2}}\,\frac{g}{\sin(k_m)J}\,\ket{\psi_{k_0}} \label{psik1}\,,
	\end{equation}
	where $\ket{\psi_{k_0}}$ is given by \eq\eqref{modes-2-1} under the replacement $k_m\rightarrow k_0$.
	
	The question is now whether or not, depending on the atoms' arrangement, ${\cal K}_{12}\neq 0$ (if ${\cal K}_{12}=0$ we get a master equation where not only the dissipator but also the effective Hamiltonian vanishes).
	Analogously to a continuous waveguide \cite{KockumPRL2018}, the two giant atoms can be arranged according to three possible topologies [see \fig\ref{fig-DFH-ex}(d)]: serial, braided and nested. Evidently, the {\it serial} configuration necessarily entails ${\cal K}_{12}= 0$ since the BS of one atom is fully confined between its coupling points, hence its overlap with the site state of the other emitter trivially vanishes.
	It is also clear that the {\it braided} configuration yields a non-zero interaction under the only condition that the coupling point of one emitter does not lie on a node of the VDS of the other emitter. The effective coupling strength in this configuration is worked out as [\cf\eq\eqref{psik1}]
	\begin{eqnarray}
		{\cal K}_{12}={\cal K}_{21}= \frac{2g^2}{v} \sin (k_0 x_{21})\,,\label{k12-prime}
	\end{eqnarray}
	where we defined the photon group velocity $v=d\omega/dk=2J\sin k_0$ [\cf\eq\eqref{omegak}].
	
	We are left with the analysis of the {\it nested} configuration [see \fig\ref{fig-DFH-ex}(d)]. In this case, evidenty, the BS of atom 2 has {\it zero} amplitude on both atom-1 coupling points, hence ${\cal K}_{21}={\cal K}_{12}\propto \langle \chi_1 |\psi_{\rm BS}^2{\rangle}=0$ resulting in a null DFH. It is interesting to point out that the BS of atom 1 generally does have {\it non-zero} amplitude on every coupling point of atom 2. However, these amplitudes sum to zero so that the overlap between $|\psi_{\rm BS}^1{\rangle}$ and $\ket{\chi_2}$ still vanishes
	\begin{equation}
		{\cal K}_{12}\propto \langle \chi_2 |\psi_{\rm BS}^1{\rangle}	\propto \left[\sin (k_0 x_{21})+\sin (k_0 x_{22})\right]=0 \label{eq12}\,, 
	\end{equation}
	in agreement with the general constraint ${\cal K}_{12}={\cal K}_{21}^*$ [see \eq\eqref{recipro}]. To derive \eq\eqref{eq12} we used the VDS condition for atom 2 [second identity of \eq\eqref{nup}]. Thus we conclude that, like the serial one, the nested configuration also yields a null DFH.

	These results are fully in line with those predicted for a continuous waveguide through fully different approaches \cite{KockumPRL2018, carollo2020mechanism} (these can be retrieved by turning the array into a continuous waveguide through linearization of the dispersion law (see \eg \rref\cite{LeonfortePRL2021}). Remarkably, this unifies the physical mechanism behind DFHs of giant atoms in a waveguide continuum with the BS-mediated picture that is usually applied to DFHs in bandgaps.

	\subsection{Sufficient condition for zero interaction}\label{cond}
	
	In general, the interaction strength for a pair of atoms $j$ and $j'$ vanishes whenever the BS of {\it one} atom ($j$ and/or $j'$) has zero amplitude on {\it every} coupling point of the other atom, \ie when $\langle x_{j\ell}|\Psi^{j'}_{\rm BS}\rangle=0$ for any $\ell$ and/or $\langle x_{j'\ell}|\Psi^{j}_{\rm BS}\rangle=0$ for any $\ell$. In the {\it serial} configuration of the setup in \fig\ref{fig-DFH-ex}(d) both identities hold, whereas in the {\it nested} geometry only one holds but the interaction strength vanishes anyway.
	\\
	\\
	\subsection{Square lattice}
	
	We finally study the case of a pair of giant atoms with ${\cal N}=4$ coupled to a homogeneous square lattice as in \fig\ref{fig-DFH-ex}(e) (braided arrangement) where each emitter's coupling points are arranged analogously to
	\fig\ref{fig-vds-ex}(d). Since there is only one coupling point of emitter 2 (1) which overlaps the BS seeded by emitter 1 (2), a DFH arises with interaction strength given by [\cf\eqs\eqref{Heff2}, \eqref{psi0} and \eqref{psi02}] ${\cal K}_{12}={\cal K}_{21}={\bar g}\langle \chi_2 | \psi_{\rm BS}^1\rangle={g^2}/{J}$ \footnote{The occurence of such a DFH was briefly mentioned in \rref\cite{gonzalez2019engineering}.}.

	The above emitters' arrangement can be considered a 2D extension of the {\it braided} configuration of Section \ref{dis-sec} and \fig\ref{fig-DFH-ex}(d). Likewise, we can define a {\it nested} geometry as shown in \fig\ref{fig-DFH-ex}(e) (bottom panel), in which case we get ${\cal K}_{12}={\cal K}_{21}=0$ due to condition \ref{cond} since the BS of the internal emitter has zero amplitude on every coupling point of the external one.
	
	\section{Decoherence-free Hamiltonian in 2D \texorpdfstring{\\}{} with two coupling points }\label{sec-2cps}
	
	In contrast to the 1D waveguide of \fig\ref{fig-DFH-ex}(c) featuring giant atoms with two coupling points, all the previous 2D instances of DFHs [\cf\figs \ref{fig-DFH-ex}(a), (b), (c), (e)] employed giant atoms with at least {\it three} coupling points. In each case, the VDS was confined within a region having three or four coupling points as vertexes. One may wonder whether three is a general lower bound on the number of coupling points in order to realise decoherence-free interactions in a 2D lattice. It turns out that this is not the case, which we show through the following counterexample in 2D using giant atoms with only two coupling points. This counterexample is again constructed by taking advantage of the VDS picture.

\begin{figure*}[t!]
		\begin{center}
			\includegraphics[width=1.\textwidth]{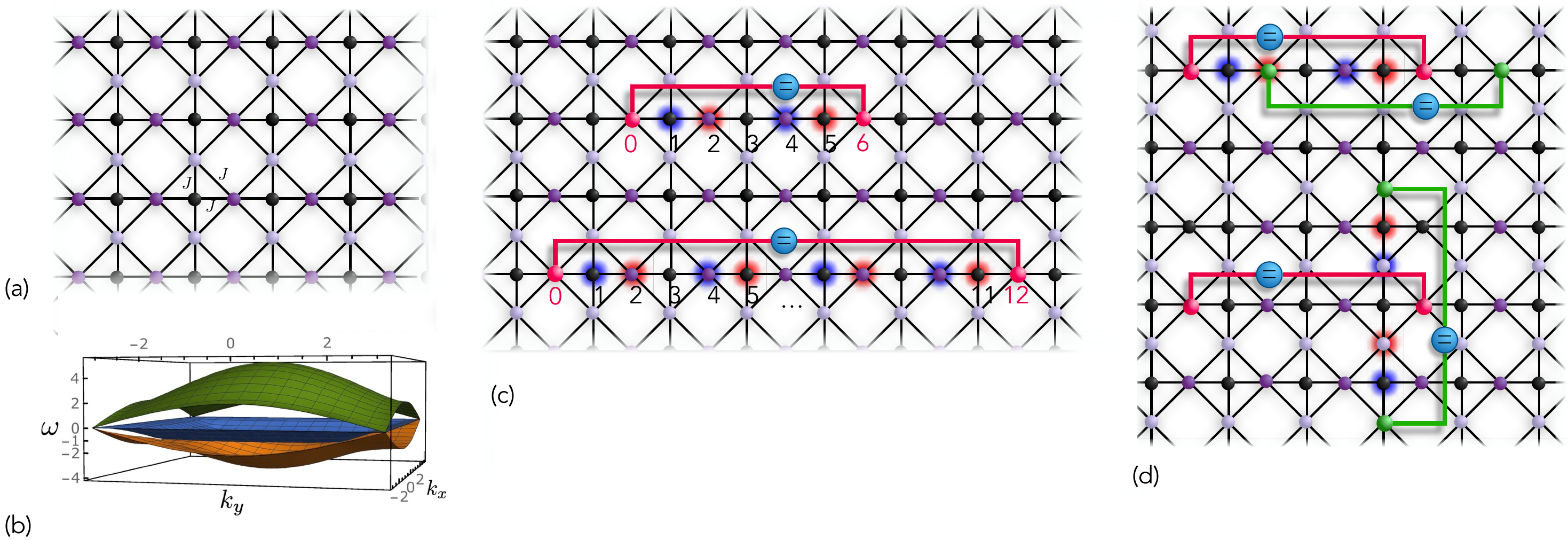}
			\caption{Giant atoms with ${\cal N}=2$ in a photonic 2D Lieb lattice with added next-nearest-neighbour hoppings. (a) Lattice structure, where the three sublattices are highlighted. All photon hopping rates (diagonal, horizontal and diagonal) are equal to $J$. (b) Dispersion laws of the three photonic bands (frequency $\omega$ in units of $J$). The bands touch one another at the corners of the first Brillouin zone. (c) Top: VDS seeded by a giant atom with two coupling points (cavities 0 and 6) coupled to the ends of a five-cavity string (cavities 1-5). The VDS has non-zero amplitude only on red and blue cavities, on which it takes uniform modulus and phase 0 ($\pi$) on red (blue) cavities.
				Bottom: giant atom coupled to an eleven-cavity string and seeded VDS. (d) Top: non-trivial DFH with two giant atoms of the same geometry (horizontal in this case). Bottom: instance of trivial DFH with giant atoms of different geometry (one horizontal one vertical). There is no way to arrange for the VDS of one atom to overlap the coupling point of the other atom.} \label{fig-lieb}
		\end{center}
	\end{figure*}

	Consider the three-partite 2D lattice sketched in \fig \ref{fig-lieb}(a). 
	This is a generalization of the standard Lieb lattice \cite{weeks2010topological} with added next-nearest-neighbour hopping rates \cite{beugeling2012topological}, where all hopping rates -- both nearest-neghbour and next-nearest-neghbour, have the same value $J$. 
	We set to zero the frequency of each cavity. As shown in \fig \ref{fig-lieb}(b), the bare lattice shows up three bands which touch one another at the four corners of the first Brillouin zone in a way that no bandgap occur. Like graphene and square lattice (see \eg \rrefs\cite{Gonzalez-Tudela2017b,gonzalez2018exotic}), depending on its frequency the decay dynamics of a normal atom can be non-Markovian even if it is tuned within the photonic continuum. In particular, this happens for $\omega_{0}=-J$ in which case the emitter undergoes a dynamics very similar to vacuum Rabi oscillations (not shown here) which can be attributed to the central photonic band in \fig \ref{fig-lieb}(b) whose dispersion law is almost flat close to this energy. In contrast, we will show next that a giant atom of the same frequency and with only two coupling points can seed a VDS.
	
	\subsection{VDS with two coupling points}
	
	Consider the linear array of $N_a=5$ cavities skecthed on top of \fig \ref{fig-lieb}(c). It is easy to see that the free Hamiltonian of this homogeneous five-cavity array admits in particular the eigenstate [\cf \fig \ref{fig-lieb}(c) for the definition of cavity labels]
	\begin{equation}
		\ket{\psi_a}=\frac{1}{2}\left(\ket{1}-\ket{2}+\ket{4}-\ket{5}\right)\,\label{psia}
	\end{equation}
	with energy $\omega_a=-J$ (note that there is a node at cavity 3). One can check by inspection that, due to destructive interference, this state decouples from each of the three cavities located right on top of 1, 3 and 5, respectively, as well as from the corresponding three cavities on the bottom . On the other hand, the remaining two lattice cavities directly linked to the array, which are labeled by 0 and 6 in \fig\ref{fig-lieb}(c), couple to state $\ket{\psi_a}$ only through their antisymmetric superposition $\ket{{\chi}}=\frac{1}{\sqrt{2}}\, (\ket{0}{-}\ket{6})$ with strength $\langle \chi|H_B| \psi_{a}\rangle=J/\sqrt{2}$. This is because the symmetric superposition of $\ket{0}$ and $\ket{6}$ decouples from state \eqref{psia} due to the opposite phases of amplitudes $\langle 1\ket{\psi_a}$ and $\langle 5\ket{\psi_a}$. Accordingly, if we couple a giant atom to cavities 0 and 6 with strengths $g_0=-g_6=g$ (\ie there are only {\it two} coupling points with a $\pi$ phase shift) and tune its frequency to $\omega_{0}=\omega_{a}=-J$, a VDS \eqref{dressed} exists with $\ket{\psi_{\rm VDS}}=\ket{\psi_a}$ whose photonic wave function reads [\cf\eq\eqref{BS-weak2-vds}]
	\begin{equation}
		\ket{\psi_{\rm BS}}=\frac{2 g}{J} \ket{\psi_{0}}\label{psi03}\,. %\frac{\sqrt{2} g}{J/\sqrt{2}} \ket{\psi_{0}}
	\end{equation}
	(in this case $\bar g=\sqrt{2} g$). In contrast to a normal atom of the same frequency, therefore, such giant atom will not decay (for weak coupling).

	\subsection{Decoherence-free effective Hamiltonian}
	
	Following the same general scheme as in the previous instances, such VDS will ensue a non-trivial decoherence-free effective Hamiltonian in the presence of two giant atoms of this form arranged \eg as in the top of \fig \ref{fig-lieb}(d) (notice that each giant atom  is coupled to a cavity where the VDS seeded by the other one has non-zero amplitude). The resulting interaction strength is worked out as [\cf\eq\eqref{Heff2} and \eqref{psi03}]
	\begin{equation}
		{\cal K}_{12}={\cal K}_{21}=-\frac{g^2}{J}% \frac{g}{2} \frac{2g}{J} %\bar{g} \braket{ \chi_1 | \psi_{\rm BS}^{2}}
	\end{equation}
	
	Observe that the wavefunction of such class of VDSs can have either horizontal geometry [as \eg in \fig\ref{fig-lieb}(c)] or vertical geometry. Interestingly, the effective Hamiltonian of two giant atoms can be non-trivial only when both giant atoms are arranged in a vertical configuration [see top of \fig \ref{fig-lieb}(d)] or both in a horizontal one. Indeed, if their configurations are different there is no way to arrange for the VDS of one atom to overlap the coupling point of the other atom [see bottom of \fig \ref{fig-lieb}(d)], entailing ${\cal K}_{12}={\cal K}_{21}=0$: the two atoms will be free from decoherence but they will not interact with one another whatsoever.

	\subsection{General case}
	
	Similarly to states \eqref{psi0} in the square lattice, the present VDSs are also scalable in size (which in this case is the length of the cavity array sandwiched between the two coupling points). Indeed, by adding to cavities 1-5 in the top of \fig \ref{fig-lieb}(c) a string of {\it six} nearest-neighbour cavities (no matter on which of the two ends of the five-cavity array), one ends up with an eleven-cavity block [see bottom of \fig \ref{fig-lieb}(c)] which again admits an eigenstate of energy $\omega_a=-J$ with wavefunction
	\begin{equation}
		\ket{\psi_a}=\frac{\ket{1}{-}\ket{2}{+}\ket{4}{-}\ket{5}{+}\ket{7}{-}\ket{8}{+}\ket{10}{-}\ket{11}}{2\sqrt{2}}\,\label{psia2}
	\end{equation}
	(nodes on cavities 3, 6, 9)
	This state has properties analogous to state \eqref{psia}: it decouples from the seven cavities right on top of 1,3,4,5,7 and 9  and get coupled only to the antisymmetric superposition of cavities 0 and 12. Therefore, a VDS and a decoherence-free Hamiltonian can be obtained through an analogous reasoning. In general, one obtains a VDS of growing length by adding iteratively a six-cavity cluster (on either end). The general size of such VDS, which in fact measures the distance between the giant-atom coupling points, is therefore $N_a=5+6 \nu$ with $\nu=0,1,...$.
	
	%Interestingly, if the giant atom above is replaced by a normal atom of the same frequency $\omega_0=-J$ it can be checked that this undergoes a non-Markovian behaviour which is very similar to vacuum Rabi oscillations \footnote{Note that also in the graphene and square lattice, for the cases considered in, a normal atom (in place of the giant atom) undergoes non-Markovian decays, which are yet of a different nature than the present vacuum Rabi oscillations.}. This can be explained by observing that, despite $\omega_0=-J$ falls within a photonic continuum, the bath features an almost flat band close to this energy (see ...), which causes a cavity-QED-like behavior.

	% where the possible values of $N_a$ are $N_a=5+3\nu$ with $\nu=0,1,2,...$ (the figure shows the case $\nu=0$).

	\section{Conclusions}\label{sec-conc}
	
	In this paper, we presented a general theoretical framework for approaching giant atoms coupled to structured photonic baths, whose experimental realization is today reachable in scenarios such as circuit QED. We first showed that one can effectively describe the system as a normal atom locally coupled to a fictitious cavity, and then used this picture to work out a compact expression of the joint atom-photon Green's function fully in terms of the free bath resolvent and the single-photon state associated with the fictitious cavity ("site state"). We next derived basic properties of atom-photon bound states (BSs), both in and out of the continuum, a general multi-atom master equation and the condition for occurrence of a decoherence-free Hamiltonian (DFH). In this way, we in particular unified known DFHs occurring within bandgaps with recently discovered DFHs in the continuum (the latter being achievable only with giant atoms), by pinpointing emergence of BSs as their common key feature. It was indeed proven that the overlap of the BS of one (giant) atom onto the site state of another one measures the strength of their mutual photon-mediated interaction. An important class of BSs was next identified: vacancy-like dressed states (VDSs), one such state being tightly connected with a localized photon eigenstate of the bath Hamiltonian but with a vacancy in place of the giant-atom's site state. 
	
	%\red 
    {We note that in all of our examples we considered coupling points with uniform strengths. Relaxing this condition generally affects emergence of VDSs and ensuing DFHs since their existence rely on a destructive interference mechanism. This is explicitly illustrated in \ref{app-g1g2} in the case of giant atoms coupled to a discrete waveguide, where we show that emergence of VDS and DFH are robust to small discrepancies between the strengths of coupling points.}
	
	The BS-based framework provides a natural interpretation of DFHs of giant atoms in the continuum of a 1D waveguide, where it is known (even experimentally) that a non-zero DFH can arise only with braided arrangements of emitters, while nested and serial configurations are ineffective: to suppress dissipation, each emitter must seed a sinusoidal VDS that yet results in a non-zero interaction only provided that the VDS of one atom overlaps the site state of another one, a condition indeed ensured solely by the braided geometry.
	
	Even more importantly, the BS-based framework combined with VDS theory work as an effective general tool to predict new classes of DFHs with giant atoms (having no analogues with normal atoms). As paradigmatic examples, we showed that giant atoms with three or four coupling points coupled to the photonic analogue of graphene or a square lattice of coupled cavities can give rise to, generally two-dimensional, DFHs. Notably, these occur within the photonic continuum, including singular points where the bath density of states locally vanishes (as in graphene) or diverges (as in the square lattice). At such points, normal atoms are unable to seed strictly bound states and/or fully lack a Markovian limit. We also showed with an explicit example (extended Lieb lattice) that DFHs in a 2D photonic continuum are possible by using giant atoms with only two coupling points, such DFHs being mediated by a BS with linear geometry.
	
	Quantum optics with giant atoms is a very young research area, especially in structured photonic baths (typically lattices), where the first related studies appeared only in the last couple of years and mostly targeted specific models through ad hoc approaches. In this paper, we instead addressed the topic from a broad perspective by developing a model-independent approach and establishing a number of general properties. 
	Since giant atoms are generally complex systems (compared to point-like emitters), it is reasonable to expect that they can bring about a zoo of interesting effects, the vast majority of which being yet unexplored (for instance coupling points with non-uniform strengths and/or phases are still little studied). We thus envisage that the framework introduced here could in particular supply a versatile methodology in order to advance the field along new unexplored avenues.
	\\
	\\
	While finishing writing this manuscript, we became aware of the related \rref\cite{Soro24}, which is being made available as a preprint at the same time as our paper.

	\ack{XS acknowledges financial support from China Scholarship council (Grant No. 202208410355). We acknowledge financial support from European Union-Next Generation EU through projects: Eurostart 2022 ``Topological atom-photon interactions for quantum technologies"; PRIN 2022--PNRR no.~P202253RLY ``Harnessing topological phases for quantum technologies"; THENCE--Partenariato Esteso NQSTI--PE00000023--Spoke 2 ``Taming and harnessing decoherence in complex networks".}
	%	\begin{acknowledgments}	
		%		XS acknowledge financial support from China Scholarship council (Grant No. 202208410355). XS, AC and FC acknowledge financial support from European Union– NextGenerationEU through project PRJ-1328 ``Topological atom-photon interactions for quantum technologies" (MUR D.M. 737/2021) and project PRIN 2022-PNRR P202253RLY ``Harnessing topological phases for quantum technologies".
		%	\end{acknowledgments}
	
	\appendix

	\section{Green's function matrix element}\label{app0}
	
	The matrix element of the bath Green's function between two (generally different) site states is
	
	\begin{equation}
		\braket{\chi_{j} | G_B ( \omega_0 ^+ ) | \chi_{j'}}=\int\! { \rm d } { \omega} \frac{ \rho_{j j'} ( \omega ) }{ \omega_0 + i \varepsilon - \omega},\label{Gb2}
	\end{equation}
	with $\epsilon\rightarrow 0^+$ and $\rho_{j j'} ( \omega ) = \sum_{n,\mathbf{k}} \delta ( \omega - \omega_{n\mathbf{k}} ) \langle \chi_j \ket{\phi_{n\mathbf{k}}}\!\bra{\phi_{n\mathbf{k}}} \chi_{j'} \rangle$. Now, using that $1/y^+=1/(y+i\epsilon)={\cal P}(1/y) - i \pi \delta(y)$ (for $y$ real) we get
	\begin{equation}
		\frac{1 }{ \omega_0 + i \varepsilon - \omega }={\cal P}\left(\frac{1}{\omega_0 - \omega }\right) - i \pi \delta(\omega -\omega_0)\,,
	\end{equation}
	which replaced in \eqref{Gb2} yields
	\begin{eqnarray}
		\braket{\chi_{j} | G_B ( \omega_0 ^+ ) | \chi_{j'}}={\cal P} \int\! { \rm d } { \omega }\,\frac{ \rho_{j j'} ( \omega )  }{ \omega_0 - \omega } - i\pi \rho_{j j'} ( \omega_0 )  \,.\label{g22}
	\end{eqnarray}
	Notice that the term $\sim{\cal P}$ is generally complex; it is yet ensured to be real in the special case $\ket{\chi_{j}}=\ket{\chi_{j'}}=\ket{\chi}$ so that \eqs\eqref{ReG} and \eqref{ImG} hold.
	
	%The matrix element of the bath Green's function between two (generally different) site states is
	%{ \color{red} 
		%\begin{equation}
		%    \braket{\chi_{j} | G_B ( \omega_0 ^+ ) | \chi_{j'}}=\int\! { \rm d } { \lambda} \frac{\langle \chi_j | \hat{\rho} ( \lambda ) | \chi_{j'}\rangle}{ \omega_0 + i \varepsilon - \lambda}\label{Gb2}
		%\end{equation}
		%}
	%with $\epsilon\rightarrow 0^+$ { \color{red} and $\hat{\rho} ( \lambda ) = \sum_{n,\mathbf{k}} \delta ( \lambda - \omega_{n\mathbf{k}} ) \ket{\phi_{n\mathbf{k}}} \!\bra{\phi_{n\mathbf{k}}}$}. Now, using that $1/y^+=1/(y+i\epsilon)={\cal P}(1/y) - i \pi \delta(y)$ (for $y$ real) we get
	%{ \color{red} \begin{equation}
			%    \frac{1 }{ \omega_0 + i \varepsilon - \lambda }={\cal P}\left(\frac{1}{\omega_0 - \lambda }\right) - i \pi \delta(\lambda -\omega_0)\,,
			%\end{equation} }
			%
			%which replaced in \eqref{Gb2} yields
			%{\color{red} \begin{widetext}
					%		\begin{eqnarray}
						%	\braket{\chi_{j} | G_B ( \omega_0 ^+ ) | \chi_{j'}}={\cal P} \int\! { \rm d } { \lambda }\,\frac{\langle \chi_j | \hat{\rho} ( \lambda ) | \chi_{j'}\rangle }{ \omega_0 - \lambda } - i\pi \langle \chi_j | \hat{\rho} ( \omega_0 ) | \chi_{j'}\rangle \,.\label{g22}
						%		\end{eqnarray}
					%\end{widetext} }
					%Notice that the term $\sim{\cal P}$ is generally complex; it is yet ensured to be real in the special case $\ket{\chi_{j}}=\ket{\chi_{j'}}=\ket{\chi}$ so that \eqs\eqref{ReG} and \eqref{ImG} hold.

					\section{Derivation of the master equation}	\label{appA}

					\eq\eqref{H1-bis-2} describes a model where the $N_a$ emitters embody the open system which is coupled to the bath $B$ according to the interaction Hamiltonian
					\begin{equation}
						H_I =\sum_{j=1}^{ N_a} \bar{g } \left(b_{\chi_j}^\dag \sigma_{j -}+\Hc\right)\,,\label{Hi}
					\end{equation}
					which fulfils $\langle {\rm vac}|H_I|{\rm vac}\rangle=0$.
					
					Passing to the interaction picture such that
					\begin{equation}
						\tilde b_{\chi_j}(t)=e^{-i H_B t} b_{\chi_j}\,,\,\,\,\tilde \sigma_{j- }(t)=e^{-i\omega_{0} t} \sigma_{j -}\label{tilde},
					\end{equation} 
					and performing the usual Born-Markov approximation, when the bath is in the vacuum state $\ket{\rm vac}$ the emitters' state $\tilde \rho$ evolves in time according to (see \eg \rref\cite{carmichael2009open})
					\begin{equation}
						\dot{\tilde \rho}(t) = -\int_0 ^t { \rm d } t'\,{\rm Tr}B\left\{ \left[\tilde H_{I}(t),\left[\tilde H_{I}(t'),\tilde{\rho}(t)| {\rm vac}\rangle \langle {\rm vac}|\right]\right]\right\},\,
					\end{equation}
					with ${\rm Tr}_B$ the trace over the bath degrees of freedom. Using \eqs\eqref{Hi} and \eqref{tilde}, we explicitly get
					\begin{equation}
						\dot{\tilde\rho} = - \sum_{j,j'}  \left[ \left(\sigma_{j+}\sigma_{j'-} \tilde \rho  -  \sigma_{j'-} \tilde\rho \sigma_{j+}\right) B_{j j'}+{\rm H.c.}  \right],\label{me11}
					\end{equation}
					with
					\begin{eqnarray}
						B_{j j'} = \bar{g}^2 \int_0 ^\infty { \rm d } \tau\, e^{ i \omega_0 \tau }  \braket{{ \rm vac} | \,\tilde b_{\chi_{j}} ( t ) \tilde b^\dagger_{\chi_{j'}} ( t - \tau ) )| {\rm vac}}.\label{b1}
					\end{eqnarray}
					(where the upper integration limit was extended from $t$ to $\infty$ as the bath correlation time is very short).

					\subsubsection*{Rates \texorpdfstring{$B_{j j'}$}{B{jj'}} in terms of the bath resolvent }
					
					Using \eqref{chi2} and $e^{-i H_B \tau}\ket{\rm vac}=\ket{\rm vac}$, the integrand of \eqref{b1} can be arranged as
					\begin{eqnarray}
						e^{ i \omega_0 \tau } \braket{{ \rm vac} | \,b_{\chi_{j}} ( t ) b^\dagger_{\chi_{j'}} ( t - \tau ) )| {\rm vac}}=\braket{ \chi_{j} | e^{-i (H_B-\omega_{0}) \tau} |{\chi_{j'}} }\,.
					\end{eqnarray}
					When this is plugged in \eqref{b1} and $e^{-i (H_B-\omega_{0}) \tau}$ is expanded in terms of energies and eigenstates of $H_B$ we get
					\begin{equation}
						B_{j j'} =\bar{g}^2\! \bra{\chi_{j}} \sum_{n,\mathbf{k}}  \int_0 ^\infty \!{ \rm d } \tau\,  e^{-i (\omega_{n\mathbf{k}}-\omega_{0}) \tau}  \ket{\phi_{n\mathbf{k}}} \!\bra{\phi_{n\mathbf{k}}}\ket{\chi_{j'}}.\label{b2}
					\end{equation}
					The improper integral is worked out by adding an infinitesimally small positive imaginary part $\epsilon\rightarrow 0^+$ to $\omega_0$ as
					\begin{equation}
						\eqalign{
							\int_0 ^\infty { \rm d } \tau\,  e^{-i (\omega_{n\mathbf{k}}-\omega_{0}) \tau}& =\lim_{t\to \infty} 	\int_0 ^t { \rm d } \tau\,  e^{-i (\omega_{n\mathbf{k}}-\omega_{0}-i \epsilon) \tau}  = \\
							& = \left.  \lim_{t\to \infty}\frac{i e^{-i (\omega_{n\mathbf{k}}-\omega_{0}-i \varepsilon) \tau} }{ ( \omega_0 + i \varepsilon - \omega_{n\mathbf{k}})} \right|^0_{t} =\frac{i }{ \omega_0 + i \varepsilon - \omega_{n\mathbf{k}}} \,\,.}
					\end{equation}
					%	\begin{eqnarray}
						%		\begin{split}
							%			\int_0 ^\infty { \rm d } \tau\,  e^{-i (\omega_{n\mathbf{k}}-\omega_{0}) \tau}& =\lim_{t\to \infty} 	\int_0 ^t { \rm d } \tau\,  e^{-i (\omega_{n\mathbf{k}}-\omega_{0}-i \epsilon) \tau} \\
							%			& =\left.  \lim_{t\to \infty}\frac{i e^{-i (\omega_{n\mathbf{k}}-\omega_{0}-i \varepsilon) \tau} }{ ( \omega_0 + i \varepsilon - \omega_{n\mathbf{k}})} \right|^0_{t}\\
							%			& =\frac{i }{ \omega_0 + i \varepsilon - \omega_{n\mathbf{k}}} \,\,.
							%		\end{split}
						%	\end{eqnarray}
					
					When this is replaced back in \eqref{b2} and recalling the definition of the Green's function \eq\eqref{GBB}, we can express each rate $B_{j j'}$ fully in terms of the bath Green's function as
					\begin{equation}
						B_{j j'} = i \bar{g}^2 \braket{\chi_{j} | G_B ( \omega_0 ^+ ) | \chi_{j'}}.\label{Bdef}
					\end{equation}
					(recall that $\omega_0 ^+ =\omega_0 +i \epsilon$).
					
					\subsubsection*{Final form in terms of \texorpdfstring{$H_{\rm eff}$}{Heff} and \texorpdfstring{${\cal D}$}{D}}
					
					One can check that \eq\eqref{me11} can be arranged in the equivalent form (we also go back to the Schr\"odinger picture)
					\begin{equation}
						\dot \rho=-i [H_{\rm eff},\rho]+ {\cal D}[\rho],\label{ME12}
					\end{equation}
					with
					\begin{eqnarray}
						H_{\rm eff}&=\sum_{j,j'}(\omega_{0} \delta_{jj'}+{\cal K}_{jj'})\sigma_{j+} \sigma_{j'-}\,,\label{Heff2sm}\\
						{\cal D}[\rho]&=\sum_{j,j'} \gamma_{jj'}  \left[  \sigma_{j'-}  \rho \sigma_{j+} -\frac{1}{2}\, \{ \rho ,\sigma_{j+}  \sigma_{j'-} \} \right]\,, \label{Drho2}
					\end{eqnarray}
					where
					\begin{equation}
						\eqalign{
							{\cal K}_{jj'}= \frac{i}{2} ( B_{j j'} - B_{j' j}^*  )\,,\, \\
							\gamma_{jj'} =B_{j j'} + B^*_{j' j}\,.\label{jgam}
						}
					\end{equation}
					%\begin{equation}
					%	\begin{split}
						%		\dot{\rho} =& -i  \left[\sum_{j,j'}  \tfrac{-i}{2} (B_{j j'} - B_{j' j}^* )\sigma_{j+} \sigma_{j'-}  , \rho\right]\\
						%		&+\sum_{j,j'} ( B_{j j'} + B^*_{j' j})\left(  \sigma_{j'-} \rho\sigma_{ j+}-  \tfrac{1}{2}\sigma_{ j+} \sigma_{j'-} \rho -  \tfrac{1}{2} \rho \sigma_{ j+} \sigma_{j'-} \right) 
						%	\end{split}
					%\end{equation}
					This matches precisely master equation \eqref{ME}.

					Finally, by plugging \eqref{g22} in \eqref{Bdef} and then the latter in \eq\eqref{jgam} we end up with \eqs\eqref{jj1} and \eqref{gg1}.
					
					{%\color{red}
                    \section{Non-uniform coupling strengths for a discrete waveguide and $\cal{N}$=2}\label{app-g1g2}
					
					Here, we consider again giant atoms for ${\cal N}=2$ in a discrete waveguide [\cf Sections \ref{sec-wav}  and \ref{dis-sec}] but now allow for coupling points of generally different strengths. 
					
					\subsection{One giant atom}\label{sec-1-at}
					
					For a single giant atom coupled to cavities $n=0$ and $n=d$ respectively with strengths $g_1$ and $g_2$, the site state reads [\cf\eq\eqref{chi-wav}]
					\begin{equation}
						\ket{\chi}=\alpha_1\ket{0}+\alpha_2  \ket{d},\,\,\label{chi12}
					\end{equation} 
					with [\cf\eqs\eqref{achi} and \eqref{chi2}] $\alpha_\ell=g_\ell/\bar{g}$ for $\ell=1,2$ and ${\bar g}=\sqrt{|g_1|^2+|g_2|^2}$.  For simplicity, we consider the case that $g_\ell>0$ (the generalization to complex coupling strengths is straightforward). Accordingly, we conveniently express $\alpha_1$ and $\alpha_2$ as
					\begin{equation}
						\alpha_1=\cos \vartheta\,,\,\,\alpha_2=\sin \vartheta,\label{alpha12}
					\end{equation}
					with $0\le \vartheta\le \pi/2$. Uniform couplings are thus retrieved for $\vartheta=\pi/4$.
					
					Based on Sections	\ref{Bs-sec} and \ref{sec-ME}, the key quantity to work out is $\braket{ \chi |  G_B ( \omega_0 ^+ )| \chi}$. To carry out this task, we recall  [\cf Section \ref{sec-wav}] that the bath Hamiltonian reads $H_B=-J \sum_ n |n\rangle\!\langle n+1|+{\rm H.c.}$ The matrix representation of the corresponding 
					bath Green's function within the band then is given by \cite{Economou2006}
					%%\begin{equation}
					\begin{equation}
						\braket{ n | G_B ( \omega^+ ) | n' } =-\frac{i}{2 J \sqrt{1 - \left( \frac{\omega}{2 J} \right)^2}} \left( - \frac{\omega}{2 J} + i \sqrt{1 - \left( \frac{\omega}{2 J} \right)^2} \right)^{|n- n'|}.\,\,\,\,\label{green-wav}
					\end{equation}
					%\end{equation}
					The spectrum of $H_B$ is given by $\omega=-2J \cos k$ with $-\pi\le k\le \pi$ (we assume to be in the thermodynamic limit, hence $k$ is a continuous variable).
					Defining wave vector $k_0$ through $\omega_{0}=-2J \cos k_0$, we get 
					\begin{equation}
						\braket{n| G_B ( \omega_0^+ ) |n'} =-\frac{i}{2 J\sin{k_0}}  \,e^{i k_0 |n - n'|}=-\frac{i}{v}  \,e^{i k_0 |n - n'|}\,,\label{Gb11}
					\end{equation}
					where in the last identity we introduced the photon group velocity $v=d\omega/dk=2J\sin k_0$.
					Using \eq\eqref{Gb11} together with \eqs\eqref{chi12} and \eqref{alpha12} we thus find
					\begin{eqnarray}
						\braket{ \chi |  {G}_B ( \omega_0 ^+ )| \chi} &= -i\frac{1}{v} \left[ 1 + 2  \alpha_1 \alpha_2  e^{i k_0 d }\right]= -i\frac{1}{v} \left[ 1 +  \sin (2\vartheta)  e^{i k_0 d }\right]\,, 
					\end{eqnarray}
					whose real and imaginary parts are given by
					\begin{eqnarray}
						{\rm Re } \braket{ \chi |  {G}_B ( \omega_0 ^+ )| \chi}  = \frac{1}{v}\sin{(2 \vartheta)} \sin{(k_0 d) }\,,\\
						{\rm Im } \braket{ \chi |  {G}_B ( \omega_0 ^+ )| \chi} = - \frac{1}{v} \left[ 1 + \sin{(2 \vartheta)}  \cos{ (k_0 d) }\right]\,.\label{im-eq-dwg}
					\end{eqnarray}
					For $g_1=g_2=g$ (entailing $\vartheta=\pi/4$) and $k_0 d=(2\nu+1)\pi$, both the real and imaginary parts vanish and we retrieve the VDS in Section \ref{sec-wav}. Notice that the imaginary term can only vanish for $\sin{ (2 \vartheta )} = 1$ and $\cos (k_0 d)=-1$, which shows that no BS in the continuum is possible when $g_1\neq g_2$. Thus having uniform couplings is an essential requirement. When this condition does not occur the atom decays. Indeed, using \eq \eqref{gam} (for $j=j'=1$) combined with \eq\eqref{im-eq-dwg} the emitter's decay rate is computed as
					\begin{eqnarray}
						\gamma = -2 \bar{g}^2 	{\rm Im } \braket{ \chi |  {G}_B ( \omega_0 ^+ )| \chi}= \frac{ 2\bar{g}^2}{ v} \left[ 1 + \sin{(2 \vartheta)}  \cos{ (k_0 d) }\right]\,.\label{gam1}%= \frac{ \bar{g}^2}{ J  |\!\sin{k_0|}} \left[ 1 + \sin{(2 \vartheta)}  \cos{ (k_0 d) }\right]\label{gam1}.
					\end{eqnarray}
					For uniform couplings (\ie for $\vartheta=\pi/4$) and $k_0 d=(2\nu+1)\pi$, we get $\gamma=0$ as expected due to the emergence of the VDS. Relaxing such conditions will instead cause the atom to decay. Notice however that $\vartheta=\pi/4$ and $k_0 d=(2\nu+1)\pi$ are stationary points of the sine and cosine functions appearing in \eqref{gam1}, implying that the atom-photon BS is stable against small deviations from such ideal parameters.

					\subsection{Two giant atoms}
					We next consider two giant atoms, in which case the respective site states read
					\begin{eqnarray}
						\ket{\chi_1 } = \alpha_1 \ket{ 0 } + \alpha_2 \ket{ d }\,,\,\,\ket{ \chi_2 } = \alpha_1 \ket{ x_{21} } + \alpha_2 \ket{ x_{21} + d }\,,\label{eq-chi12}
					\end{eqnarray} 
					where, in line with Section \ref{dis-sec}, atom 1 has coupling points $x_{11}=0$ and $x_{12}=d$ while atom 2 is coupled to cavities $x_{21}$ and $x_{22}$ (coefficients $\alpha_1$ and $\alpha_2$ are analogous to \ref{sec-1-at}). We will focus on the braided configuration [see \fig\ref{fig-DFH-ex}(d)] such that $0<x_{21}<d$ and $x_{22}>d$.
					
					Under weak coupling, the dynamics of the two atoms is governed by master equation \eqref{ME} with [\cf\eqs \eqref{Js} and \eqref{gam}]
					\begin{eqnarray}
						\mathcal{K}_{12} = \mathcal{K}_{21}^* =\bar{g}^2 \frac{ \braket{ \chi_1 |  {G}_B ( \omega_0 ^+ )| \chi_2} + \braket{ \chi_2 |  {G}_B ( \omega_0 ^+ )| \chi_1}^* }{2}\,,\label{k12-1} \\
						\gamma_{12} =\gamma_{21}^*= i \bar{g}^2 \left(  \braket{ \chi_1 |  {G}_B ( \omega_0 ^+ )| \chi_2} - \braket{ \chi_2 |  {G}_B ( \omega_0 ^+ )| \chi_1}^* \right).\label{gam12-1}
					\end{eqnarray}\,
					Using \eqs\eqref{green-wav} and \eqref{eq-chi12} we have
					\begin{eqnarray}
						\braket{ \chi_1 |  {G}_B ( \omega_0 ^+ )| \chi_2} &= -\frac{i}{v} \,\xi \,,\,\,\,\,\braket{ \chi_2 |  {G}_B ( \omega_0 ^+ )| \chi_1}^* &= \frac{i}{v}\,\xi^*,\,
					\end{eqnarray}
					with $\xi=e^{i k_0 x_{21}} + \alpha_1  \alpha_2 (e^{ i k_0 (x_{21} + d)} +  e^{i k_0 (d - x_{21})})$. Plugging these into \eqs\eqref{k12-1} and \eqref{gam12-1} and with the help of \eq\eqref{alpha12} we thus find
					\begin{eqnarray}
						\mathcal{K}_{12} = \frac{\bar{g}^2}{v} \left[ \sin{(k_0 x_{21})} + \sin{(2 \vartheta)} \sin{(k_0 d)} \cos{(k_0 x_{21})} \right]\,, \\
						\gamma_{12} = \frac{2\bar{g}^2}{v}\cos{(k_0 x_{21})} \left[ 1 +  \sin{(2 \vartheta)}  \cos{ (k_0 d )} \right] \,.
					\end{eqnarray}
					In line with \ref{sec-1-at}, when the VDS exists, \ie for $\vartheta=\pi/4$ and $k_0 d=(2\nu+1)\pi$, we get a decoherence-free dynamics since $\gamma_{12}=0$ with the atom-atom effective coupling strength given by $\mathcal{K}_{12}=\frac{2g^2}{v}\sin{(k_0 x_{21})} $ in agreement with \eq\eqref{k12-prime}. This phenomenon is robust to deviations from the ideal VDS condition, such as a small discrepancy between $g_1$ and $g_2$, for reasons analogous to those discussed at the end of Section \ref{sec-1-at}.
					}
					
					\section*{References}
					
					\bibliography{WQEDdefects,WQEDedge}
					\bibliographystyle{iopart-num}
					%\bibliographystyle{unsrt}
					%\bibliography{WQEDedge}
					%\bibliographystyle{apsrev4-2}
					%\bibliographystyle{unsrt}

				\end{document}